\documentclass[12pt]{reportj}
\usepackage{deluxetablej}
\usepackage{hyperref} 
\usepackage{amsmath} 
\usepackage[]{natbib}
\makeatletter
\def\@to{to}
\makeatother
\usepackage{times}
\usepackage{graphicx}
\usepackage{xcolor}
\usepackage{tocloft}
\usepackage{fancyheadings}
\emergencystretch=\maxdimen
\usepackage{datetime}
\newdateformat{ddmonthyyyy}{\THEDAY\ \monthname[\THEMONTH]\ \THEYEAR}

\renewcommand\thesection{\arabic{section}}
\renewcommand\thesubsection{\thesection.\arabic{subsection}}

\def\appsection#1{%
  \setcounter{subsection}{0}%
  \refstepcounter{section}%
  \section*{\hbox to \hsize{\large\bf Appendix \Alph{section}. #1\hfill}}%
  \addcontentsline{toc}{section}{Appendix \Alph{section}. #1}}

\def\ssection#1{\setcounter{subsection}{0} \refstepcounter{section} \section*{\hbox to \hsize{\large\bf \arabic{section}. #1\hfill }}\label{sec} \addcontentsline{toc}{section}{\arabic{section}. #1}}
\def\ssubsection#1{\setcounter{subsubsection}{0} \refstepcounter{subsection}\subsection*{\hbox to \hsize{\normalsize\bfseries\itshape \arabic{section}.\arabic{subsection} #1\hfill}}\label{subsec} \addcontentsline{toc}{subsection}{\arabic{section}.\arabic{subsection} #1}}
\def\ssubsubsection#1{\refstepcounter{subsubsection}\subsection*{\hbox to \hsize{\normalsize\it \arabic{section}.\arabic{subsection}.\arabic{subsubsection} #1\hfill}}\label{subsubsec} \addcontentsline{toc}{subsubsection}{\arabic{section}.\arabic{subsection}.\arabic{subsubsection} #1}}

\def\ssectionstar#1{\section*{\hbox to \hsize{\large\bf #1\hfill}} \addcontentsline{toc}{section}{#1}}
\def\ssubsectionstar#1{\subsection*{\hbox to \hsize{\normalsize\bfseries\itshape #1\hfill}} \addcontentsline{toc}{subsection}{#1}}
\def\ssubsubsectionstar#1{\subsection*{\hbox to \hsize{\normalsize\it  #1\hfill}} \addcontentsline{toc}{subsection}{#1}}

\renewcommand{\cftaftertoctitle}{%
\mbox{}\hfill{\normalfont Page}}
\defcitealias{dressel2007}{STIS ISR 2007-03}
\defcitealias{Ward-Duong2022}{STIS ISR 2022-01}
\defcitealias{friedman2005}{STIS ISR 2005-03}
\defcitealias{welty2018}{STIS ISR 2018-04}
\defcitealias{welty2025}{STIS ISR 2025-01}
\defcitealias{mingozzi2026}{STIS ISR 2026-04}
\defcitealias{stisihb}{STIS Instrument Handbook}

\begin{document}

~\\

\vspace{-2.4cm}
\noindent\includegraphics*[width=0.295\linewidth]{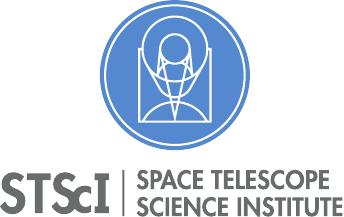}

\vspace{-0.4cm}

\begin{flushright}
    {\bf Instrument Science Report STIS 2026-03}
    
    \vspace{1.1cm}

    {\bf\Huge CCD Rotation Rates From Regular Dispersion Solution Monitoring}
    
    \rule{0.25\linewidth}{0.5pt}
    
    \vspace{0.5cm}
    
    Matilde Mingozzi$^1$, Matthew Siebert$^2$
    \linebreak
    \newline
    \footnotesize{$^1$ AURA for ESA, Space Telescope Science Institute, 3700 San Martin Drive, Baltimore, MD 21218, USA\\}
    \footnotesize{$^2$ Space Telescope Science Institute, 3700 San Martin Drive, Baltimore, MD 21218, USA\\}
    
    \vspace{0.5cm}
    
     \ddmonthyyyy\today 
\end{flushright}

\vspace{0.1cm}

\noindent\rule{\linewidth}{1.0pt}
\noindent{\bf A{\footnotesize BSTRACT}}

{\it \noindent 
It has been previously studied that the STIS CCD exhibits a slow rotation, evident from changes in spectral traces and flat-field evolution. In this Instrument Science Report, we present an independent spectroscopic analysis of the CCD rotation angle and a comparison with previous studies. We describe the methodology used to derive rotation angles and rates across different gratings and central wavelengths, and report the corresponding median rotation rates over time. The results are broadly consistent within uncertainties with earlier measurements, indicating the CCD rotation has the same impact on all the CCD gratings. We also discuss key assumptions limiting our analysis. These findings contribute to the ongoing monitoring of STIS CCD rotation and its implications for calibration accuracy.}

\vspace{-0.1cm}
\noindent\rule{\linewidth}{1.0pt}

\renewcommand{\cftaftertoctitle}{\thispagestyle{fancy}}
\tableofcontents


\newpage

\vspace{-0.3cm}
\ssection{Introduction}\label{sec:Introduction} 
The STIS CCD exhibits a slow rotation as evidenced by a change in the spectral traces \citepalias[$\sim 0.0031-0.0041$~deg/yr; ][]{dressel2007}, the evolution of the STIS CCD flat fields \citepalias[$\sim 0.0031$~deg/yr; ][]{Ward-Duong2022}, and an independent astrometric analysis of archival imaging using sources in Gaia DR2 \citep[$\sim 0.0038$~deg/yr; ][]{nguyen2021}.
Also, as reported in previous studies \citepalias{friedman2005} and confirmed by dispersion solution monitor programs \citepalias{welty2018,welty2025}, a systematic wavelength shift has been detected in spectra extracted at the CCD edges.  

A measurable rotation of the spectral traces of some CCD settings has been quantified by \citetalias{dressel2007} (i.e., G230LB/2375, G430L/4300, G750L/7751, and G750M/6768, 6581, 8561), using exposures of the standard stars taken for STIS CCD sensitivity monitoring calibration programs.
The measured rotation angle is stored in the DEGPERYR column of the one-dimensional Spectrum Trace Table (introduced in 2006; \citetalias{dressel2007}), which is used to correct the shift of the trace position on the spatial axis. 
In this ISR, we aim to independently quantify the effect of the CCD rotation on spectroscopic observations, evaluate consistency across different gratings, and compare our results with previous analyses.
In particular, we provide an independent determination of the CCD rotation angle using HITM1 lamp spectroscopic observations from the CCD dispersion solution monitoring programs\footnote{\url{https://www.stsci.edu/hst/instrumentation/stis/calibration}} spanning from Cycle 11 to Cycle 33 (approximately two decades). 
In parallel, the STIS team investigated methods to correct for the effect of CCD rotation on the dispersion direction and thereby improve the wavelength calibration accuracy across the detector, as described by \citetalias{mingozzi2026} (see also \href{https://www.stsci.edu/contents/news/stis-stans/may-2026-stan}{May 2026 STAN}).

In Section~\ref{sec:ccdrot}, we explain how we evaluated the impact of CCD rotation on the dispersion axis of HITM1 spectroscopic data. In Section~\ref{sec:deltax} and Section~\ref{sec:angles}, we describe the method used to quantify it and to derive the corresponding rotation angle and rotation rate over time for different gratings and central wavelengths. 
Finally, in Section~\ref{sec:conclusion}, we provide a brief summary of our results.


\lhead{}
\rhead{}
\cfoot{\rm {\hspace{-1.9cm} Instrument Science Report STIS 2026-03 Page \thepage}}

\section{The effect of CCD rotation on the dispersion axis}\label{sec:ccdrot}
\begin{figure}[t]
    \centering
      \includegraphics[width=1.\linewidth]{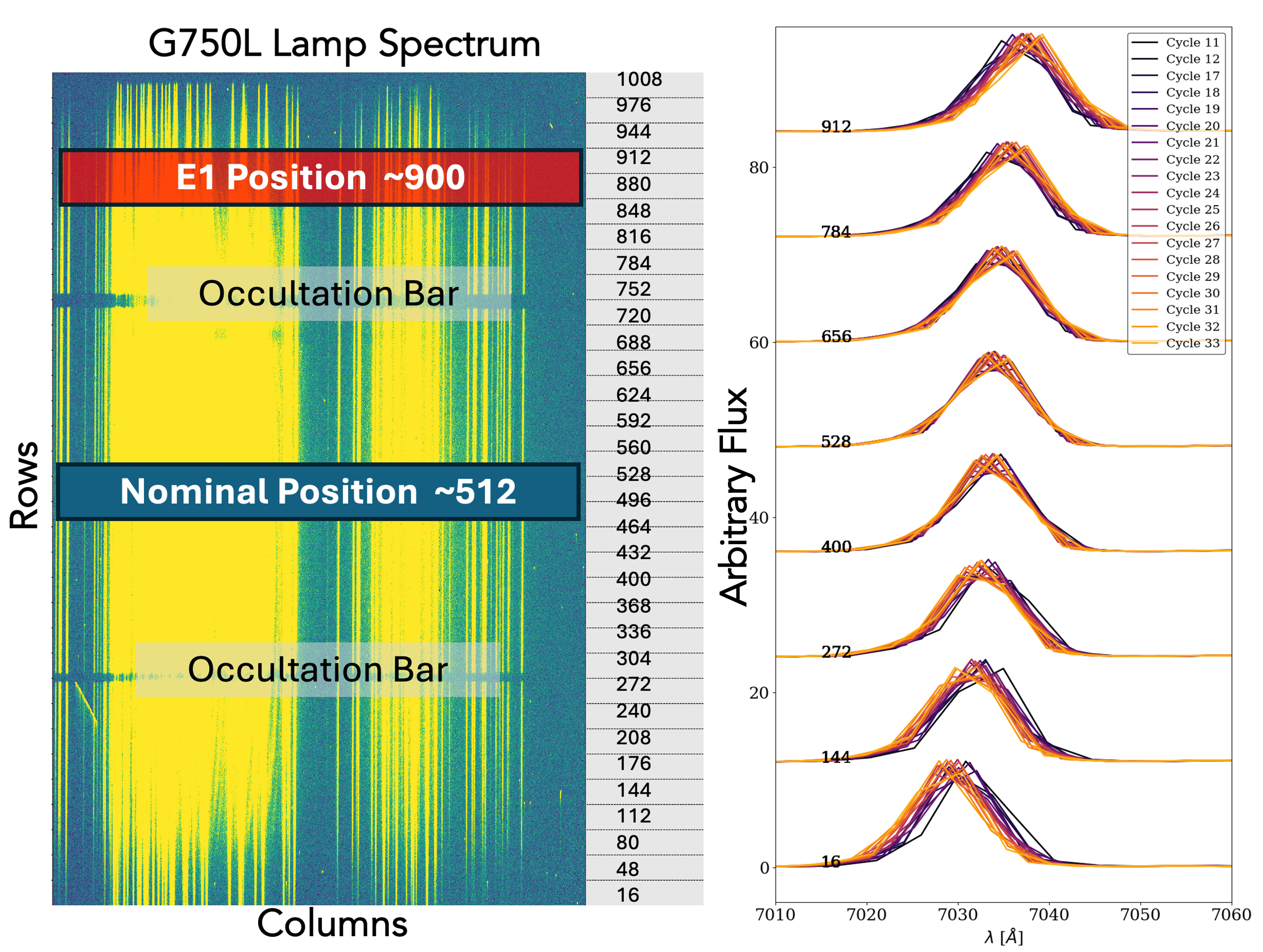}
      \caption{Example of a G750L 2D lamp spectrum from one of the dispersion solution monitor programs (left panel), from which we extracted 32 spectra for our analysis. The right panel shows examples of extracted spectra at different positions and cycles, zooming in on a single emission line to illustrate the systematic shift, which becomes more pronounced toward the edges of the CCD over time. The displayed line is redshifted (positive shift) at row 912 from Cycle 11 to Cycle 33, and blueshifted (negative shift) at position 16 over the same cycles, relative to the center of the CCD.}
         \label{fig:example1}
   \end{figure}

The left panel of Figure~\ref{fig:example1} shows an example of the 2D lamp spectrum from one of the dispersion solution monitor programs, from which we extracted 32 spectra at different positions, indicated by the gray bands (see Section~\ref{sec:deltax} for the detailed analysis explanation). 
The right panel of Figure~\ref{fig:example1} illustrates a subset of extracted spectra over time (from Cycle 11 to Cycle 33), zoomed in on a single emission line for clarity.
These datasets do not allow us to measure the effect of the rotation across the cross-dispersion or spatial axis, since the lamp illuminates the detector nearly uniformly till the edges. 
However, a clear shift along the dispersion axis is evident and can be tracked over time.
To provide an independent measurement of the rotation angle alongside the rotation trace analysis ($\Delta y$) from \citetalias{dressel2007}, we first had to determine how to estimate it from the shifts along the dispersion axis ($\Delta x$) of spectra at different CCD positions, using emission line positions as references.

\begin{figure}[!h]
    \centering
      \includegraphics[width=.85\textwidth]{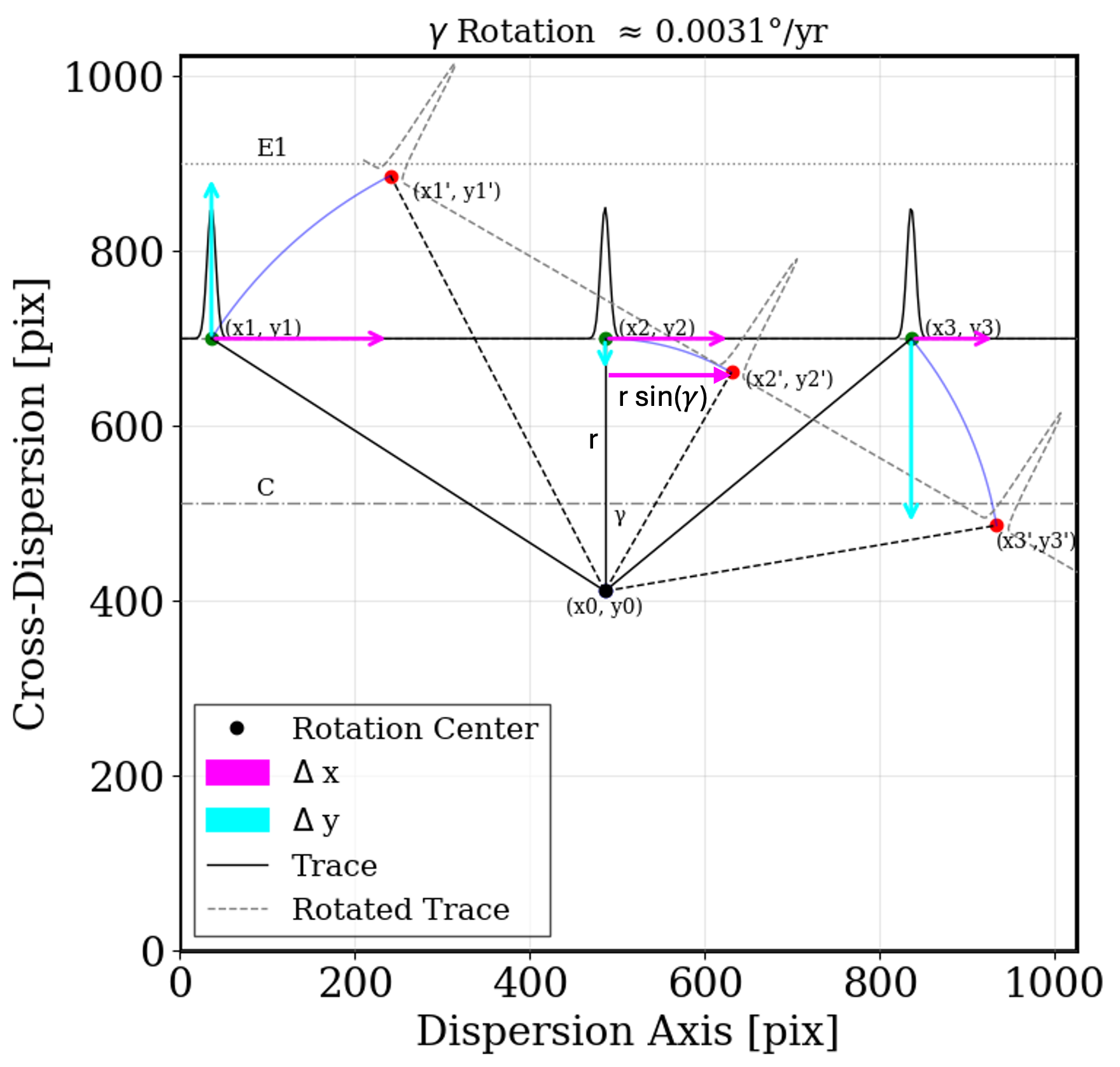}
      \caption{Simplified illustration of the effect of a rigid rotation on a spectral trace, approximated as a line on the CCD detector, with three emission lines at different positions on the dispersion axis. The dash-dotted (C) and dotted (E1) horizontal lines mark the nominal trace position (row$\sim$512) and the E1 pseudo-aperture position (row$\sim$900), respectively. The rotation angle is exaggerated to $\gamma\sim30^\circ$ for clarity. In this schematic case, both $\Delta x$ and $\Delta y$ are clearly visible, and $\Delta x$ varies along the dispersion direction. For the small rotation angle considered in this work, however, the variation of $\Delta x$ along a given detector row is negligible ($\sim10^{-3}$ pixel). The figure illustrates how the rotation angle $\gamma$ can be estimated from the measured horizontal displacement $\Delta x$ and the vertical distance from the rotation center. As shown in Appendix~\ref{sec:Appendix}, this relation can be applied across the detector to first order following the small-angle approximation. \looseness=-2}
         \label{fig:ccdrotation}
   \end{figure}

First, we adopted the clockwise rotation rate of $\sim 0.0031$~deg/yr around the center coordinates given in pixels by \citetalias[][$x_0 = 468.02$~pixel, $y_0 = 411.18$~pixel]{Ward-Duong2022}. 
We also assumed that the spectra on the CCD can be approximated as perfectly parallel lines (neglecting the small edge distortions), as shown in Fig.~\ref{fig:ccdrotation}. 
A rigid clockwise rotation of a point {\it (x,y)} on these lines by an angle $\gamma$ around the center ($x_0,y_0$) is given by:
\[
\left\{
\begin{aligned}
x' = x_0 + (x - x_0)\cos\gamma + (y - y_0)\sin\gamma \\
y' = y_0 - (x - x_0)\sin\gamma + (y - y_0)\cos\gamma
\end{aligned}
\right.
\]
where $(x', y')$ are the coordinates of the point after rotation.
With this simplification, we estimated the maximum shifts along the dispersion and cross-dispersion axes by evaluating the effect of the rotation at the four corners of the CCD detector (simply illustrated in Fig.~\ref{fig:ccdrotation}): (0,0), (1024,0), (0,1024), and (1024,1024).
Over 29 years of STIS observations (1997–2026), the shifts in both directions are clearly measurable. 
Along the dispersion axis, $\Delta x = (x' - x)$ is approximately $1$ pixel at the top corners and $-0.6$ pixels at the bottom corners.
Along the cross-dispersion axis, the shift $\Delta y = (y' - y)$ is about $0.8$ pixels on the left side and $-0.8$ pixels on the right side.
Considering the differential shifts, the variations in $\Delta x'$ across the left and right corners, and in $\Delta y'$ across the top and bottom corners, are of order $\sim 10^{-3}$ pixels. 
This shows that, although shifts along both the dispersion and cross-dispersion directions are measurable, their dependence on the columns (dispersion-axis) and rows (cross-dispersion axis), respectively, is negligible. 
This implies that, to first order, $\Delta x$ depends on the cross-dispersion coordinate $y$ and is nearly constant along the dispersion axis, while $\Delta y$ depends on the dispersion coordinate $x$ and is nearly constant along the cross-dispersion axis.

Figure~\ref{fig:ccdrotation} provides a simplified illustration of the effect of CCD rotation on spectra, showing a spectral trace with three emission lines: one near the top-left ($x_1$, $y_1$), one at the top-center ($x_2$, $y_2$), vertically aligned with the assumed rotation axis ($x_0$, $y_0$), and one near the top-right ($x_3$, $y_3$). The comparison of the solid and dashed traces offers a visual example of the rotation effect, with an exaggerated rotation angle for clarity.
The simplest case is ($x_2$, $y_2$), given that $x_2\equiv x_3$, for which the corresponding rotation angle $\gamma$ can simply be measured by applying basic trigonometric principles:
\begin{equation}\label{eq:eq1}
\gamma = \arcsin\left(\frac{\Delta x}{r}\right) \approx \frac{\Delta x}{r} \quad (\text{if } \gamma \ll 1)
\end{equation}
where $\Delta x = (x_2'-x_2)$ is the centroid shift between the two Gaussians (highlighted in magenta) and $r = y_2-y_0$ is the vertical distance from the rotation center. 
For the Gaussians located at the left and right edges of the detector in Figure~\ref{fig:ccdrotation}, the corresponding right-triangle construction cannot be used directly (see Equations~\ref{eq:gamma_scalar}--\ref{eq:r_definition} in Appendix~\ref{sec:Appendix}). However, Appendix~\ref{sec:Appendix} shows that, under the small-angle approximation, Equation~\ref{eq:eq1} can still be used to estimate the rotation angle from each measured x-displacement along the spectral trace. As illustrated in Figure~\ref{fig:ccdrotation}, the x-displacement varies with position along the dispersion axis. For the actual rotation accumulated over 29 years, however, this variation is only of order ($\sim10^{-3}$) pixels, as discussed above, and is therefore negligible for the present analysis.
In Section~\ref{sec:deltax}, we show how we measured {\it $\Delta x$} for the STIS CCD over time, evaluating the temporal shift of the emission line centroids of the HITM1 lamp, used consistently for dispersion solution monitor programs since Cycle 11 (e.g., Figure~\ref{fig:example1}). 

\section{Measurements of the shifts in time on the dispersion axis}\label{sec:deltax}
To measure the dispersion shift of spectra at vertical distance $r$ from the rotation center across time, we took into account the CCD dispersion solution monitor programs from Cycle 11 to Cycle 33 (PID: 9617, 10025, 11858, 12407, 12768, 13137, 13540, 13987, 14419, 14825, 15390, 15554, 15743, 16345, 16552, 16954, 17382, 17637, 18162), that make use of the HITM1 lamp and the G230MB, G230LB, G430M, G430L, G750M and G750L gratings at different central wavelengths.

\begin{figure}[!h]
    \centering
      \includegraphics[width=0.85\linewidth]{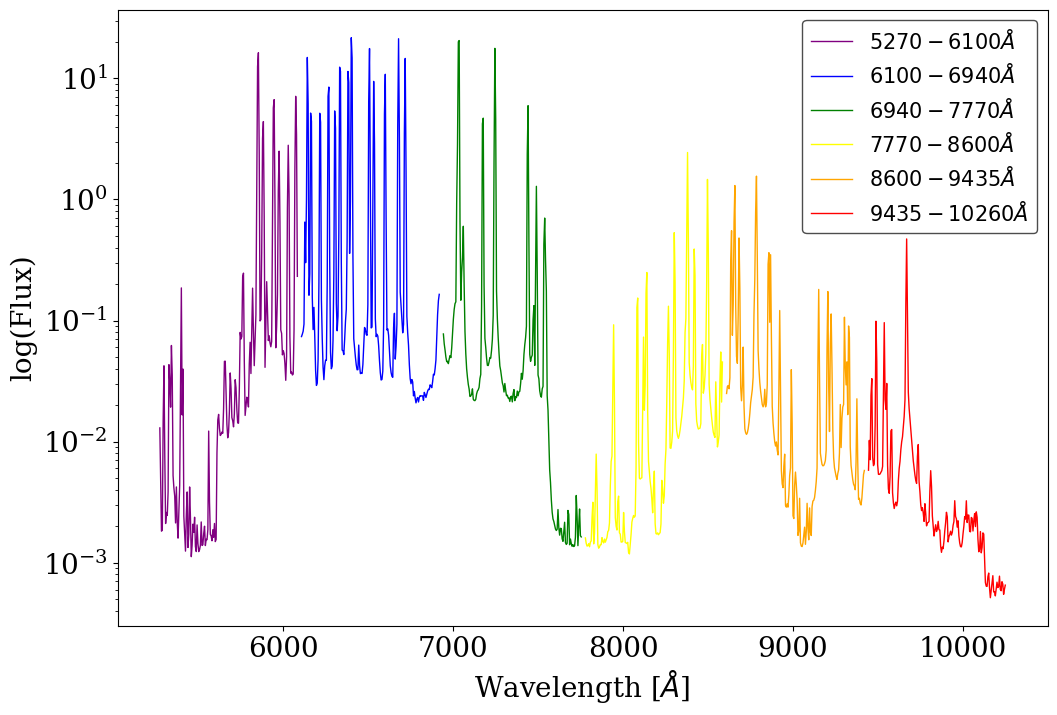}
      \caption{Example of a G750L lamp spectrum at the center of the CCD from one of the dispersion solution monitor programs, showing the total and the sub-wavelength range where the cross-correlation is performed.}
         \label{fig:example}
   \end{figure}

First, we run the \texttt{calstis} pipeline on each grating and central wavelength as described in \citetalias{friedman2005} to calibrate the data in wavelength. 
Then, we extracted from the calibrated \texttt{\_flt} lamp files 32 spectra, centred in different rows, from 16 to 1008, applying \texttt{stistools.x1d} task\footnote{To get consistent results for all the gratings we did not consider the 'DEGPERYR' information of the 1-d Spectrum Trace Table introduced after \citetalias{dressel2007} study mentioned above.}. 
This approach allows us to examine the effect of the CCD rotation along the entire length of the CCD (i.e., at different rows) as a function of time, as illustrated in Figure~\ref{fig:example1} for G750L/7751.
Finally, to measure $\Delta x$, we cross-correlated spectra extracted at the
same detector row in different cycles, using Cycle~11 as the reference epoch.
We then fitted a quadratic function to the cross-correlation curve and took the peak of the fit as the measurement of the pixel lag between each cycle and the reference cycle. 
To further test possible (despite unexpected) wavelength-dependent effects from the CCD rotation, we also divided the spectra of each grating into 6 sub-ranges (see Figure~\ref{fig:example} as an example for G750L), performing a dedicated cross-correlation across cycles for each sub-range. 
    
\begin{figure}[!h]
    \centering
      \includegraphics[width=1.\textwidth]{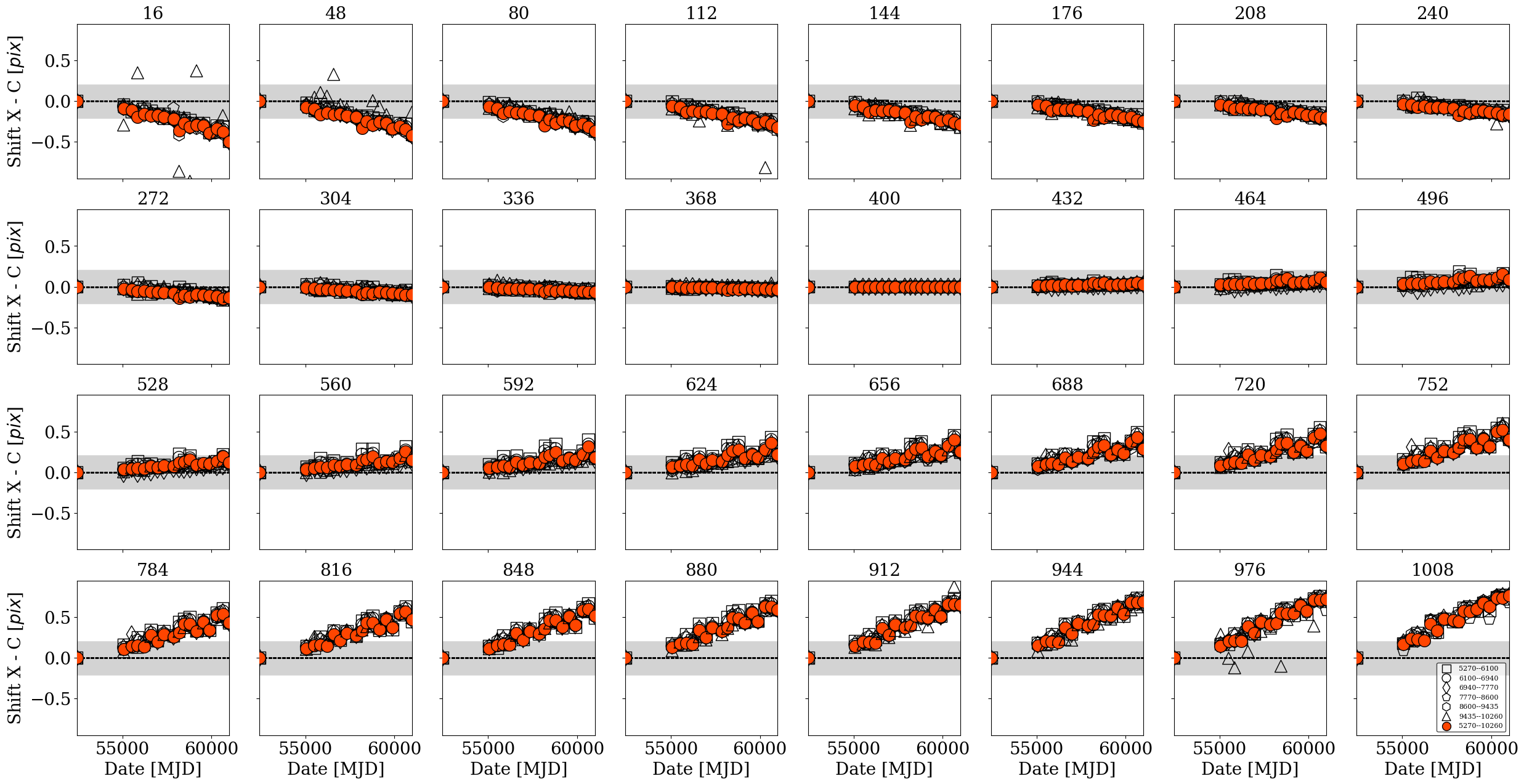}
      \caption{Shifts in pixels between the Gaussian line centroids ({\it$\Delta$x}) as a function of time for G750L/7751 (relative to Cycle 11) measured at different positions on the CCD (indicated on the label of each panel) with the cross-correlation and centred at row 400. The gray horizontal bands show the $\pm0.2$~pixels expected accuracy for the wavelength calibration. The red dots show the results obtained on the entire wavelength range, while the black points behind show the (consistent) results obtained on sub-wavelength ranges as reported in the legend. The fact that {\it$\Delta$x} is negative at the bottom of the CCD and positive at the top is in agreement with the example shown in Figure~\ref{fig:example1}.} 
         \label{fig:g750l-crosscorr}
   \end{figure}

\begin{figure}[!h]
    \centering
    \includegraphics[width=1.\textwidth]{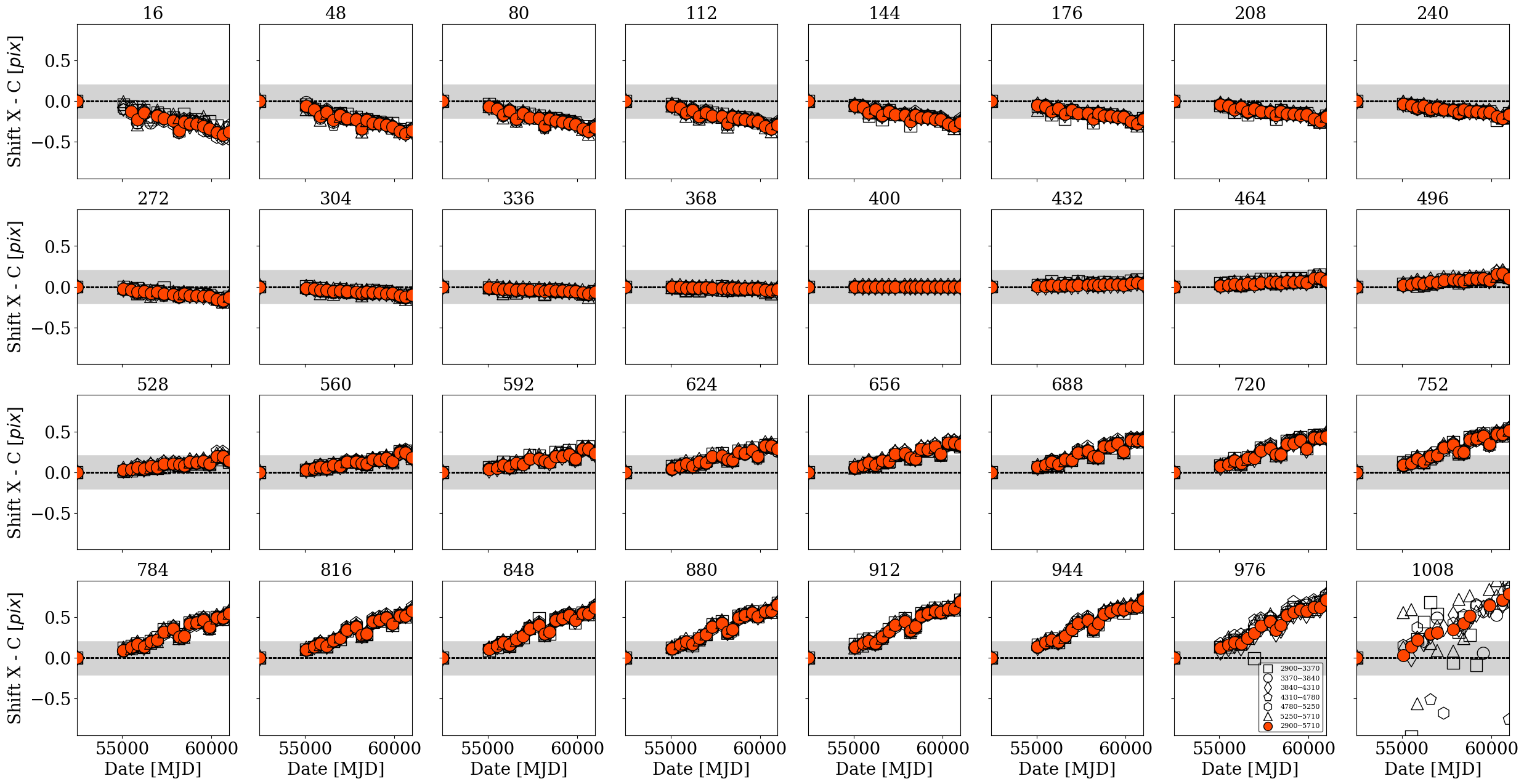}
      \caption{Same as Figure~\ref{fig:g750l-crosscorr} for G430L/4300, showing the cross-correlation results obtained on the entire wavelength range.}
         \label{fig:g430l-crosscorr}
   \end{figure}

\begin{figure}[!h]
    \centering
    \includegraphics[width=1.\textwidth]{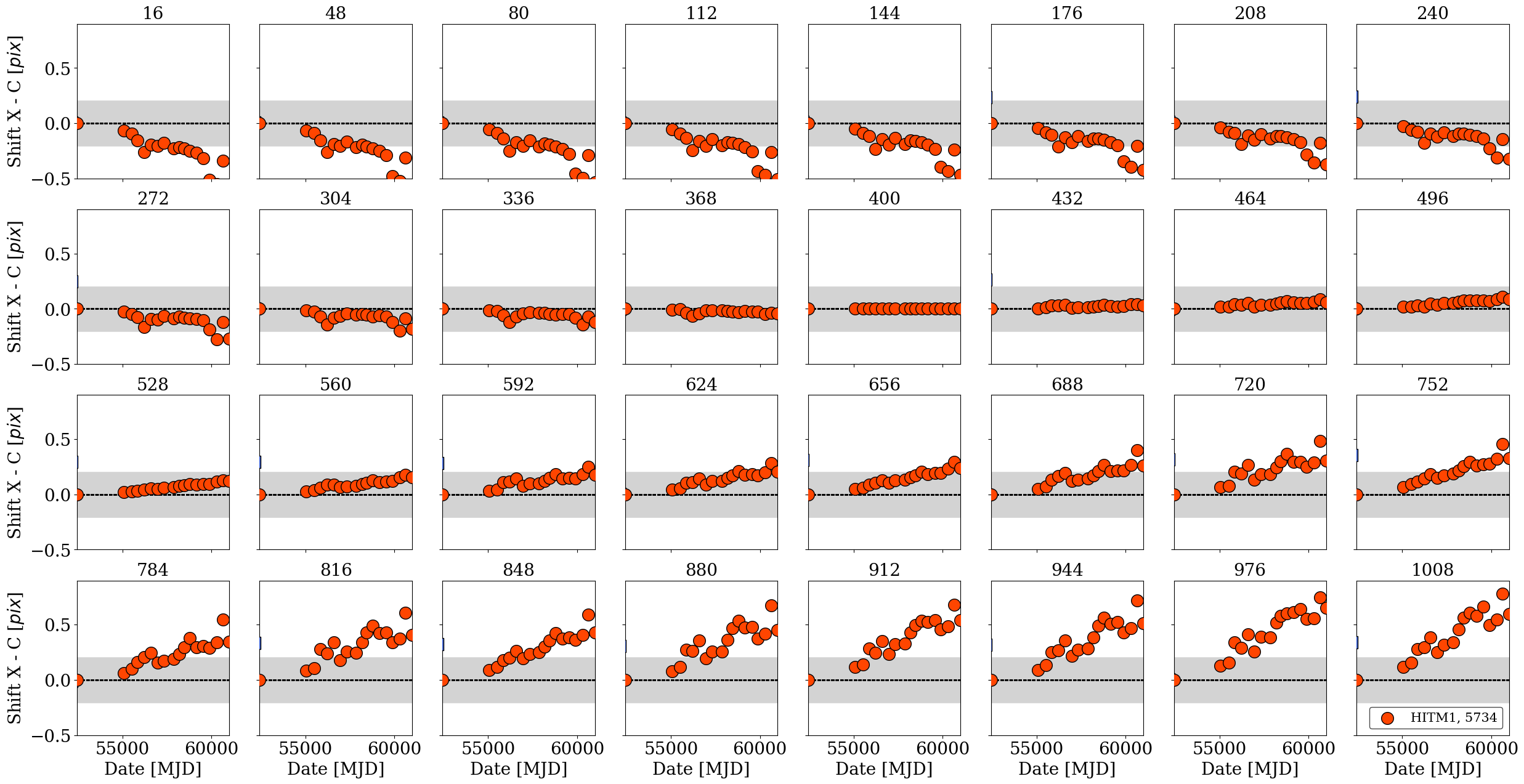}
    \includegraphics[width=1.\textwidth]{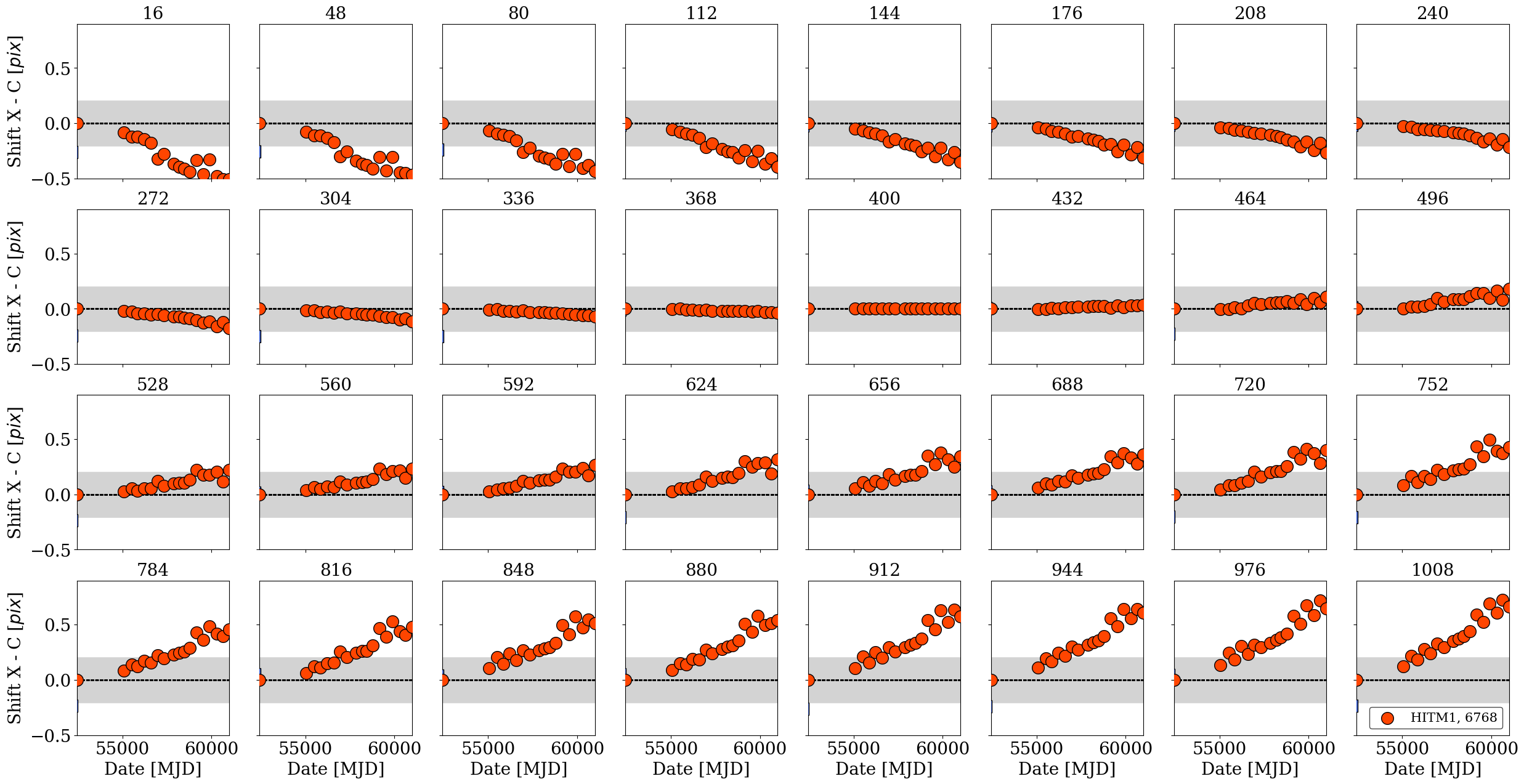}
      \caption{Same as Figure~\ref{fig:g750l-crosscorr} for G750M/5734 (top panel) and G750M/6768 (bottom panel), showing the cross-correlation results obtained on the entire wavelength range.}
         \label{fig:g750m1-crosscorr}
   \end{figure}

\begin{figure}
    \centering
    \includegraphics[width=1.\textwidth]{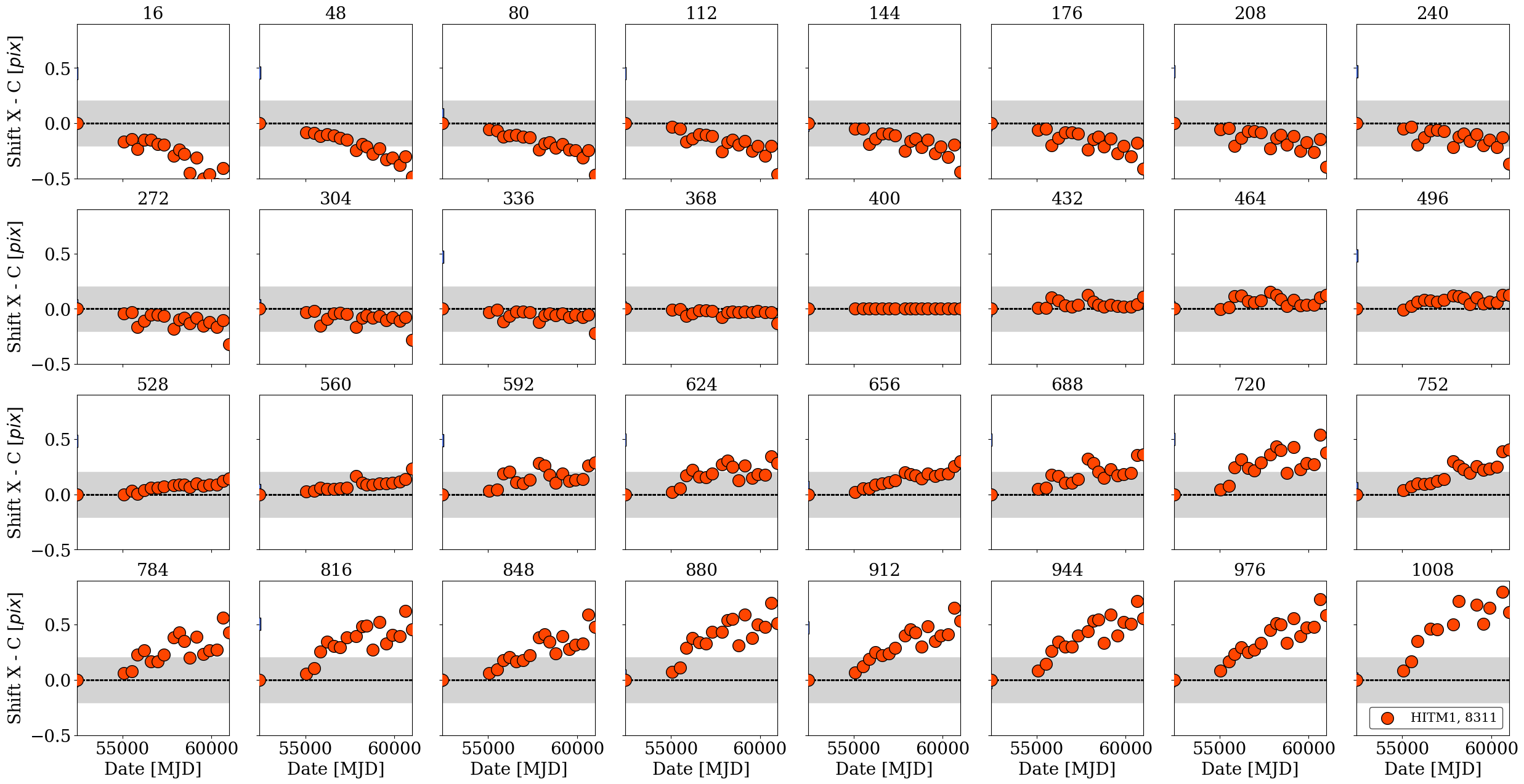}
    \includegraphics[width=1.\textwidth]{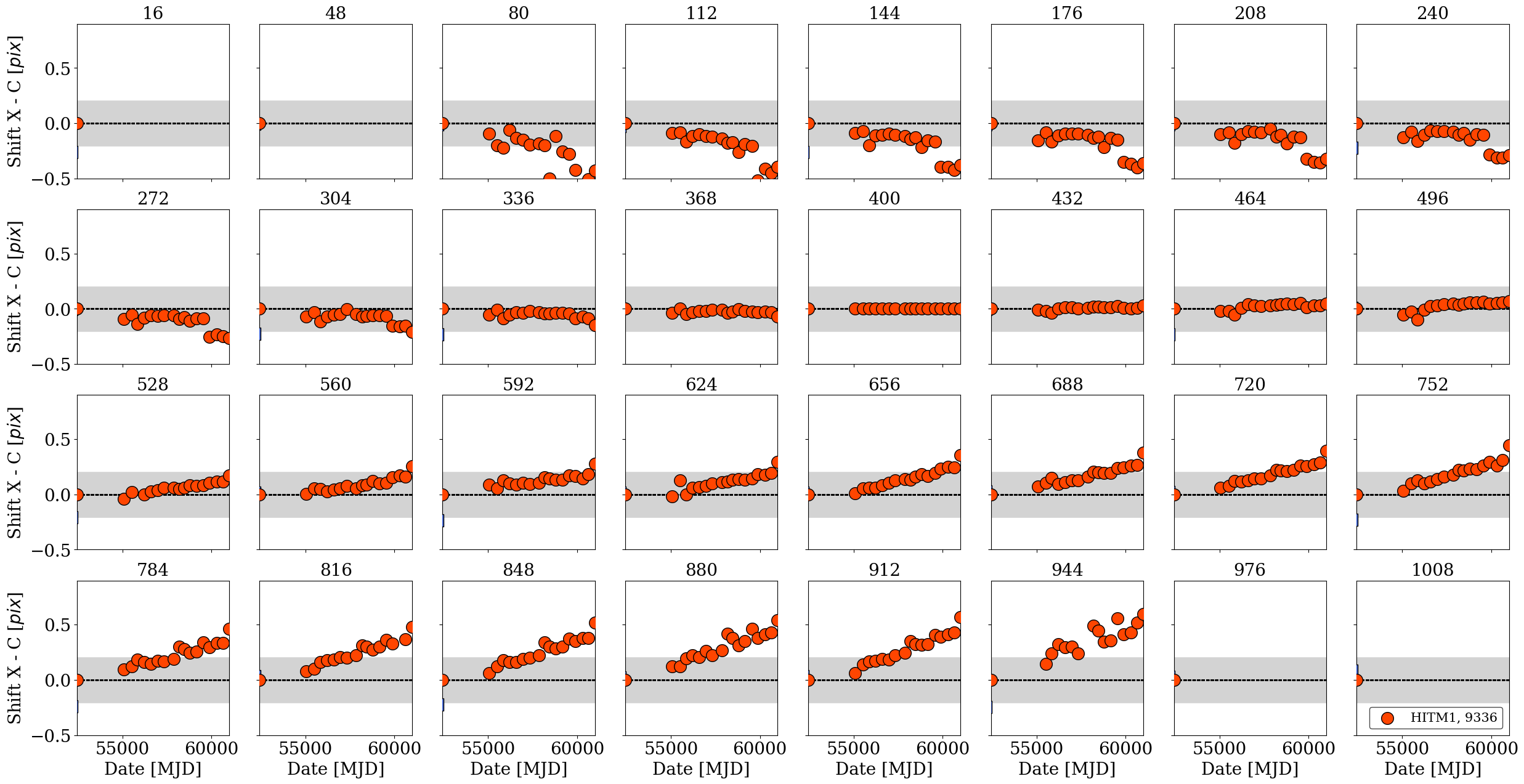}
      \caption{Same as Figure~\ref{fig:g750l-crosscorr} for G750M/8311 (top panel) and G750M/9336 (bottom panel), showing the cross-correlation results obtained on the entire wavelength range. For G750M/9336, the cross-correlation failed at the CCD edges due to the low signal-to-noise ratio in the extracted spectra in those regions.}
         \label{fig:g750m2-crosscorr}
   \end{figure}

\begin{figure}[!h]
    \centering
    \includegraphics[width=1.\textwidth]{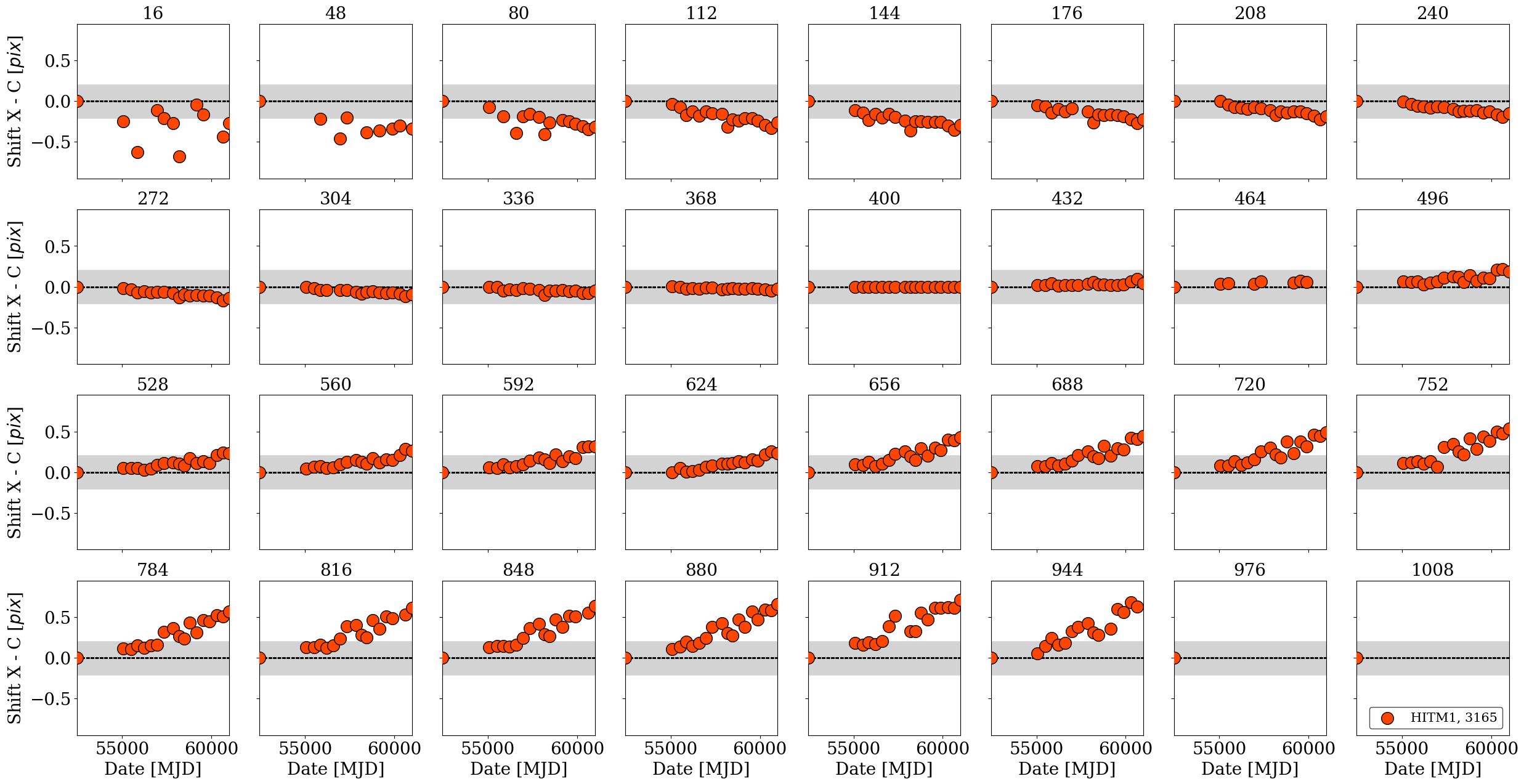}
    \includegraphics[width=1.\textwidth]{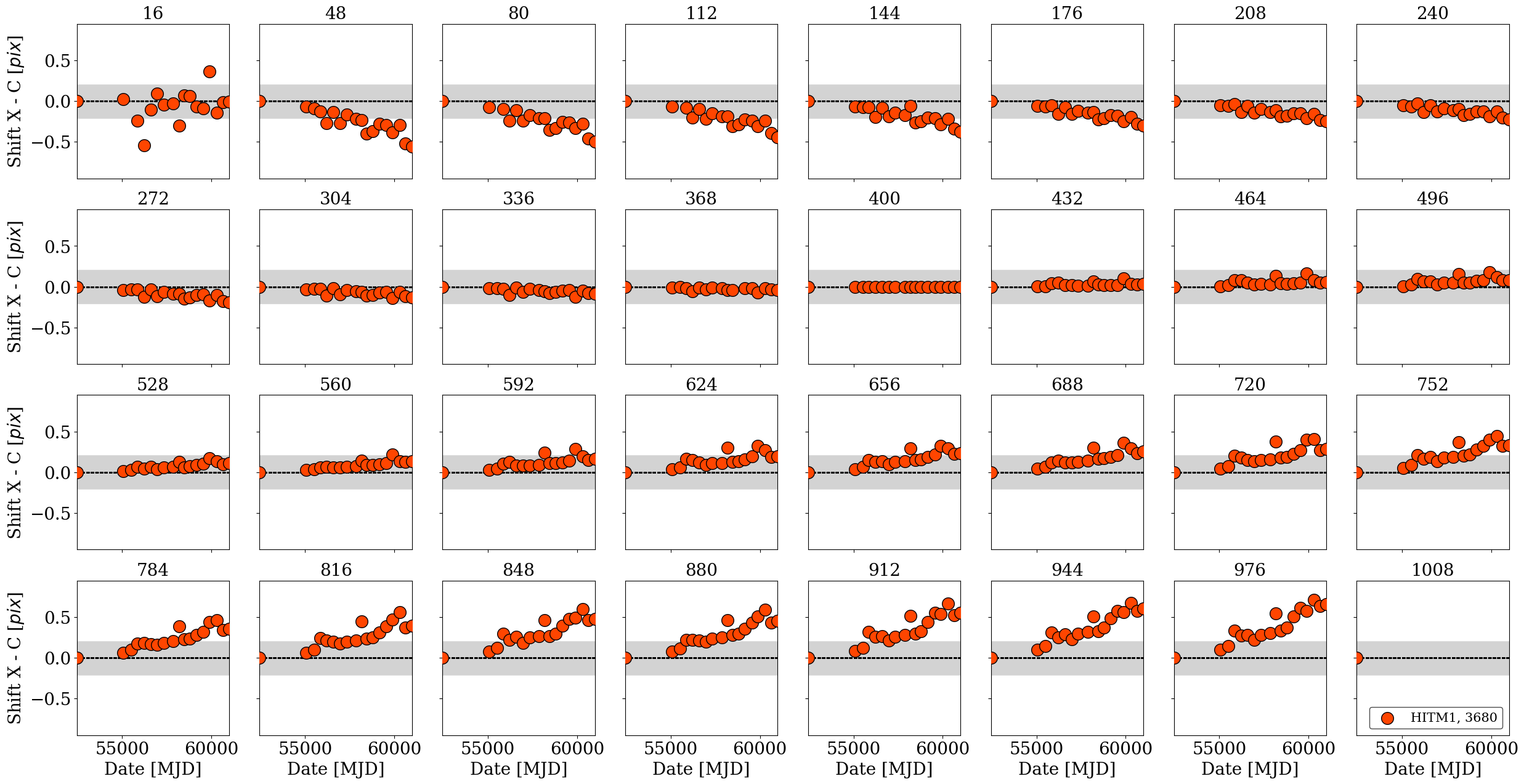}
      \caption{Same as Figure~\ref{fig:g750l-crosscorr} for G430M/3165 (top panel) and G430M/3680 (bottom panel), showing the cross-correlation results obtained on the entire wavelength range. The cross-correlation returned highly scattered values, or failed entirely, at the CCD edges due to the low signal-to-noise ratio of the extracted spectra in those regions.}
         \label{fig:g430m1-crosscorr}
   \end{figure}

\begin{figure}[!h]
    \centering
    \includegraphics[width=1.\textwidth]{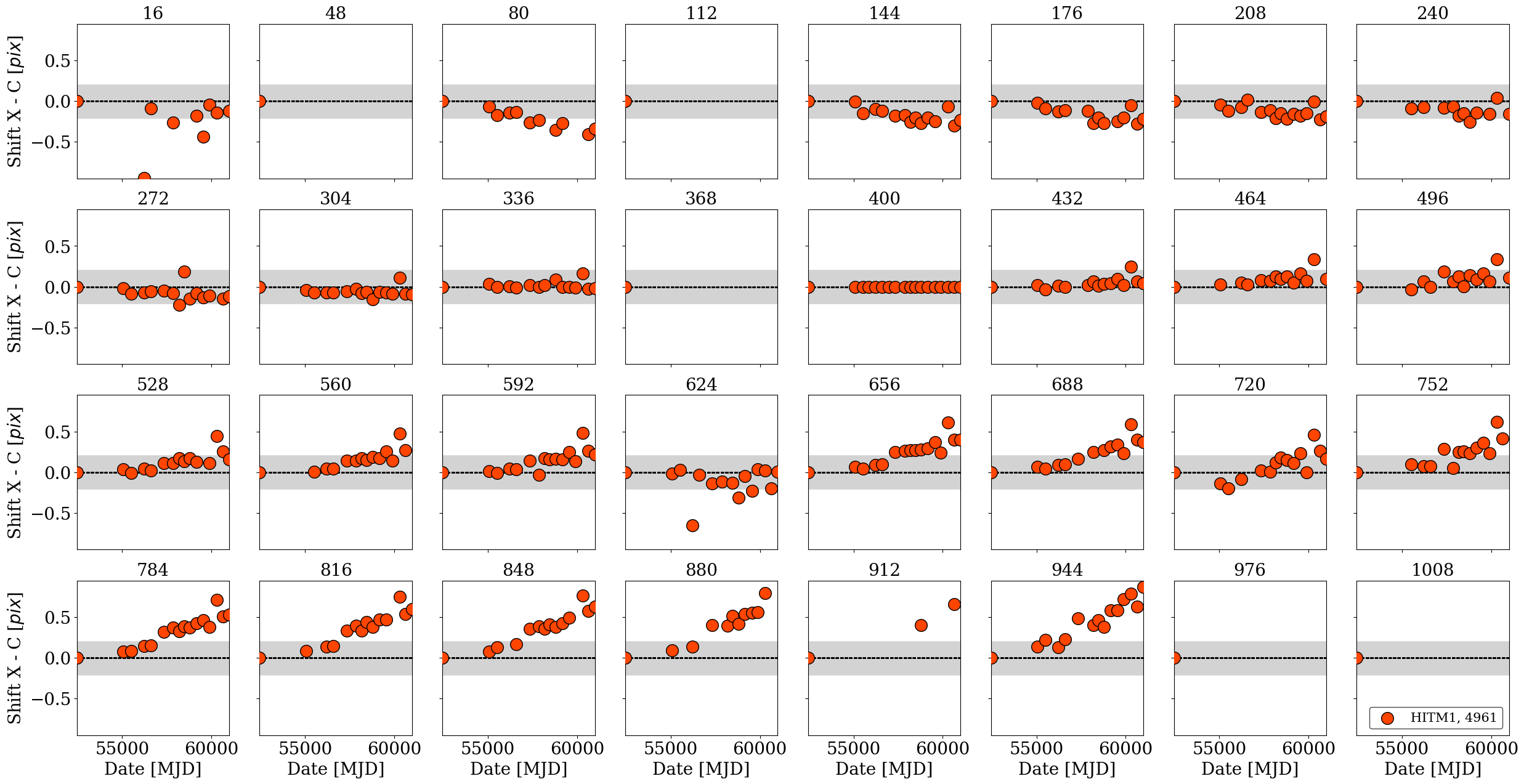}
    \includegraphics[width=1.\textwidth]{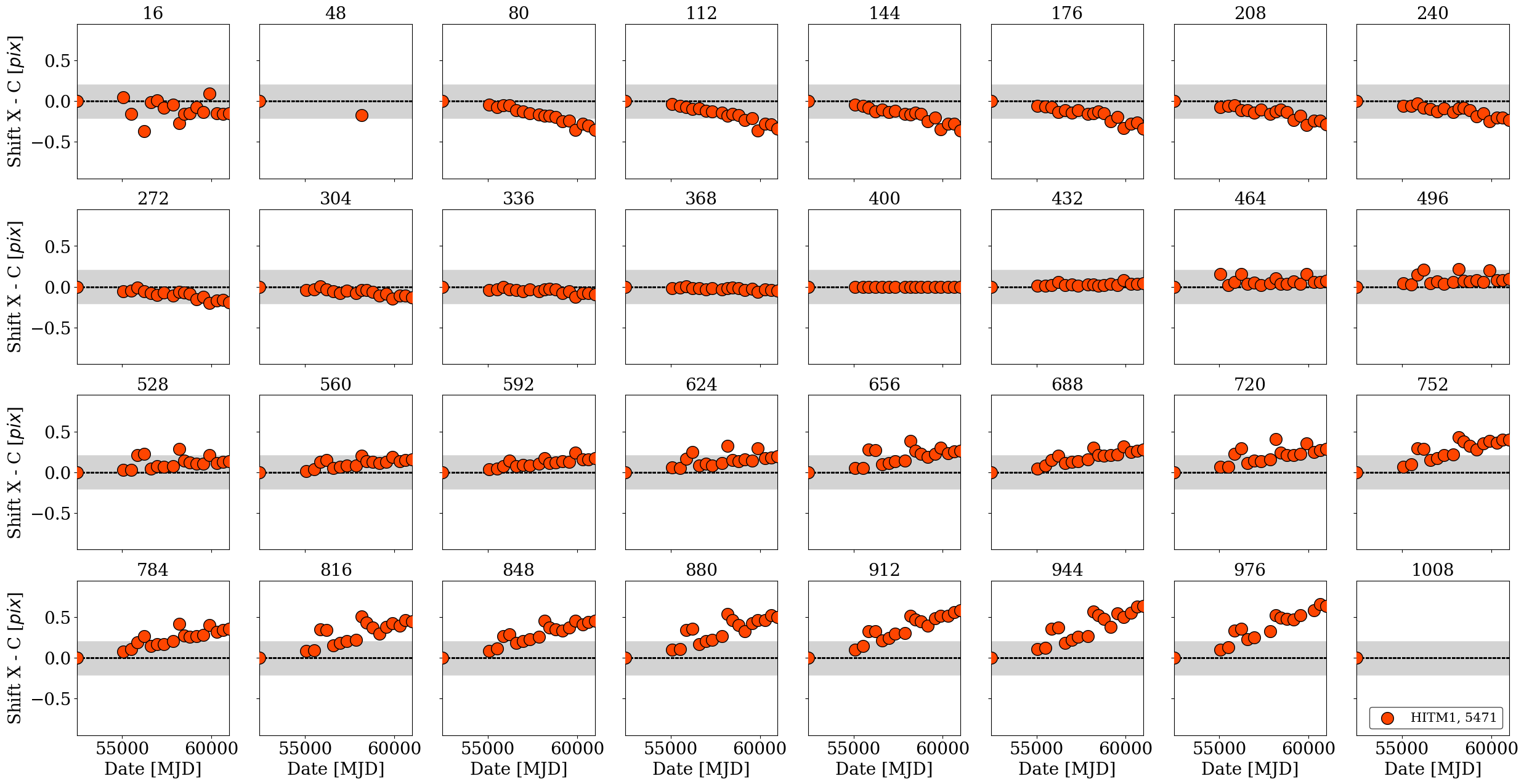}
      \caption{Same as Figure~\ref{fig:g750l-crosscorr} for G430M/4961 (top panel) and G430M/5471 (bottom panel), showing the cross-correlation results obtained on the entire wavelength range. The cross-correlation returned highly scattered values, or failed entirely, at the CCD edges due to the low signal-to-noise ratio of the extracted spectra in those regions. For G430M/4961, it also failed at row 368, likely due to a localized issue in the spectrum at that position.}
         \label{fig:g430m2-crosscorr}
   \end{figure}

\begin{figure}[!h]
    \centering
    \includegraphics[width=1.\textwidth]{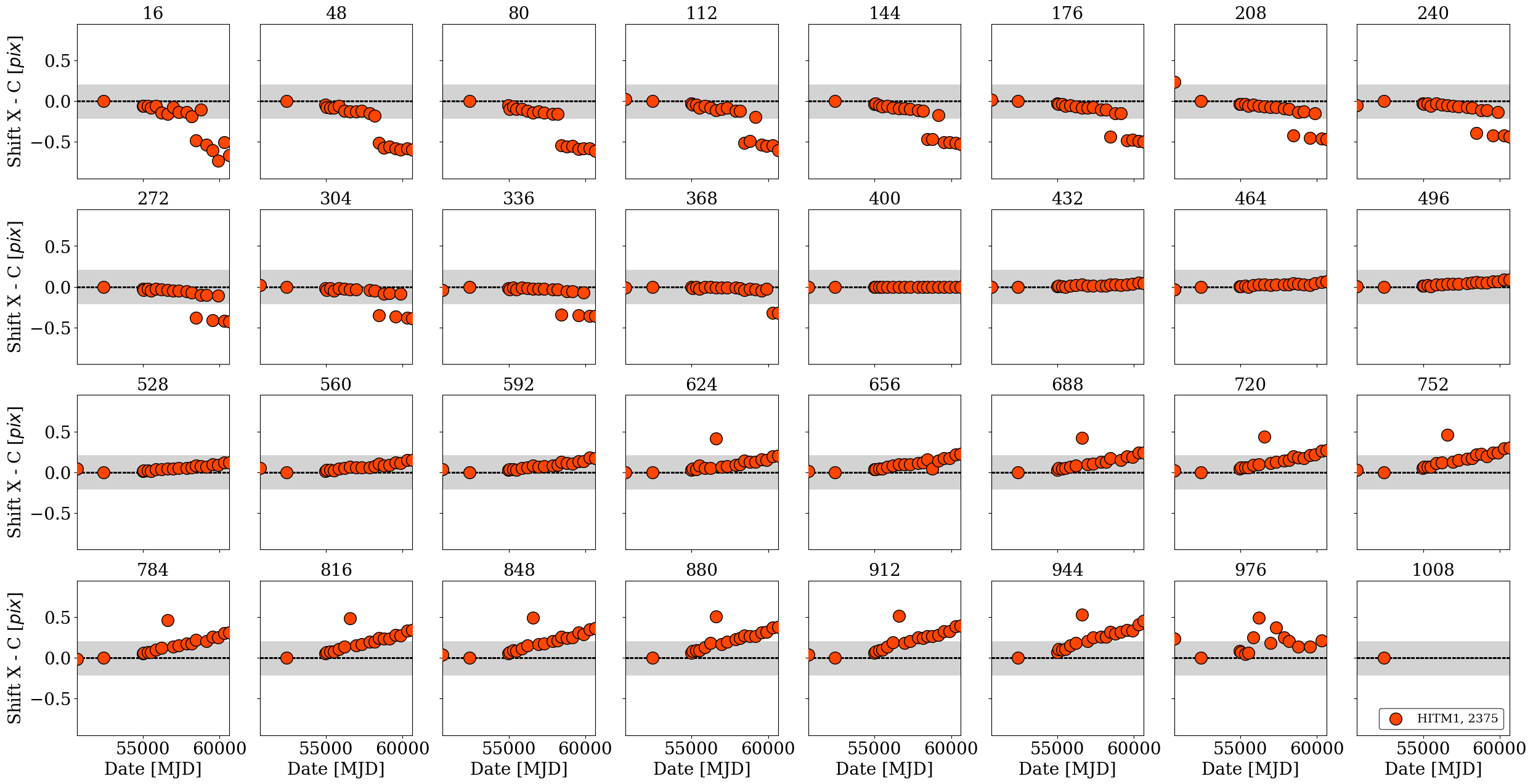}
      \caption{Same as Figure~\ref{fig:g750l-crosscorr} for G230LB/2375 showing the cross-correlation results obtained on the entire wavelength range. The cross-correlation returned highly scattered values, or failed entirely, at the CCD edges due to the low signal-to-noise ratio of the extracted spectra in those regions.}
         \label{fig:g230lb-crosscorr}
   \end{figure}

\begin{figure}[!h]
    \centering
    \includegraphics[width=.8\textwidth]{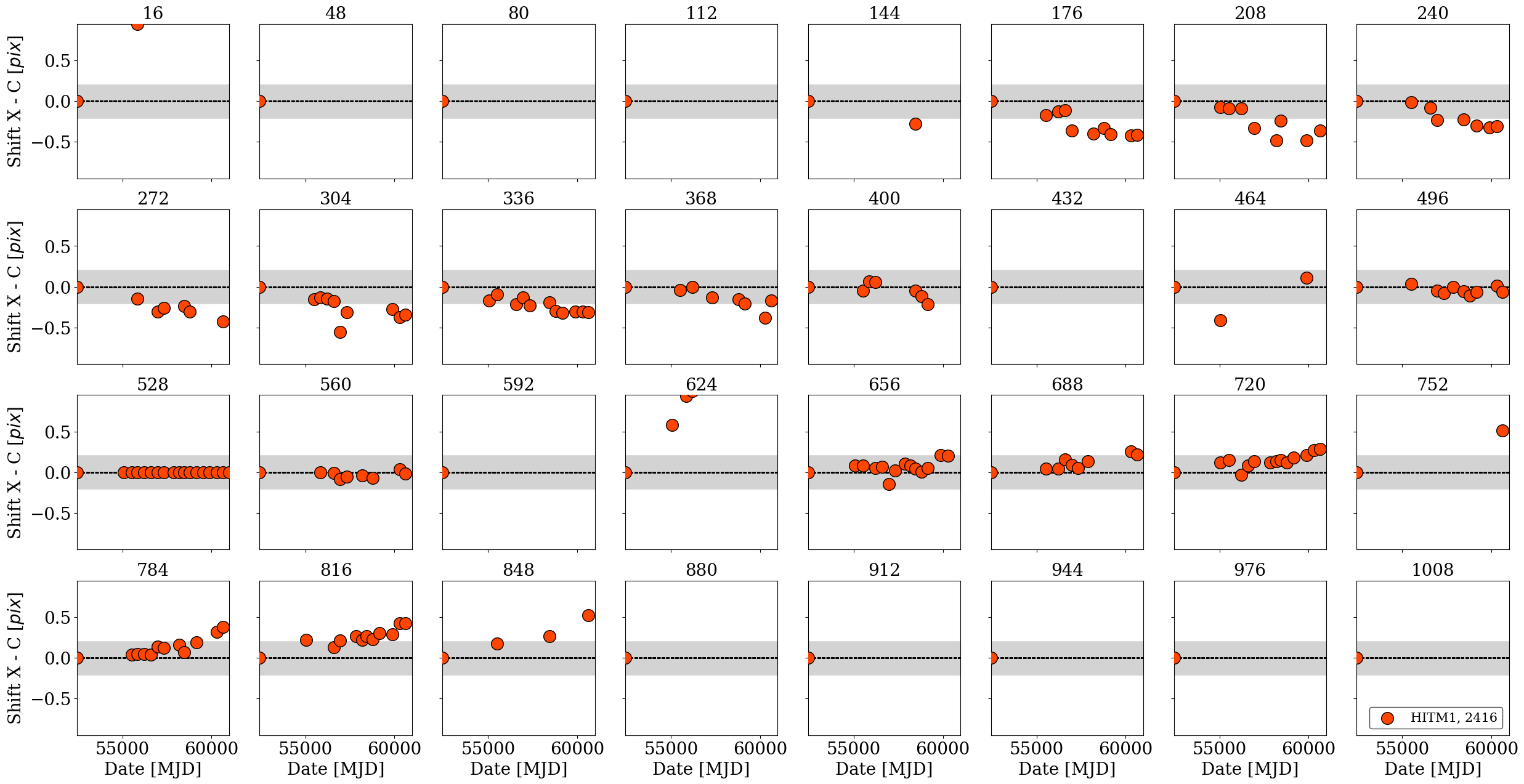}
    \includegraphics[width=.8\textwidth]{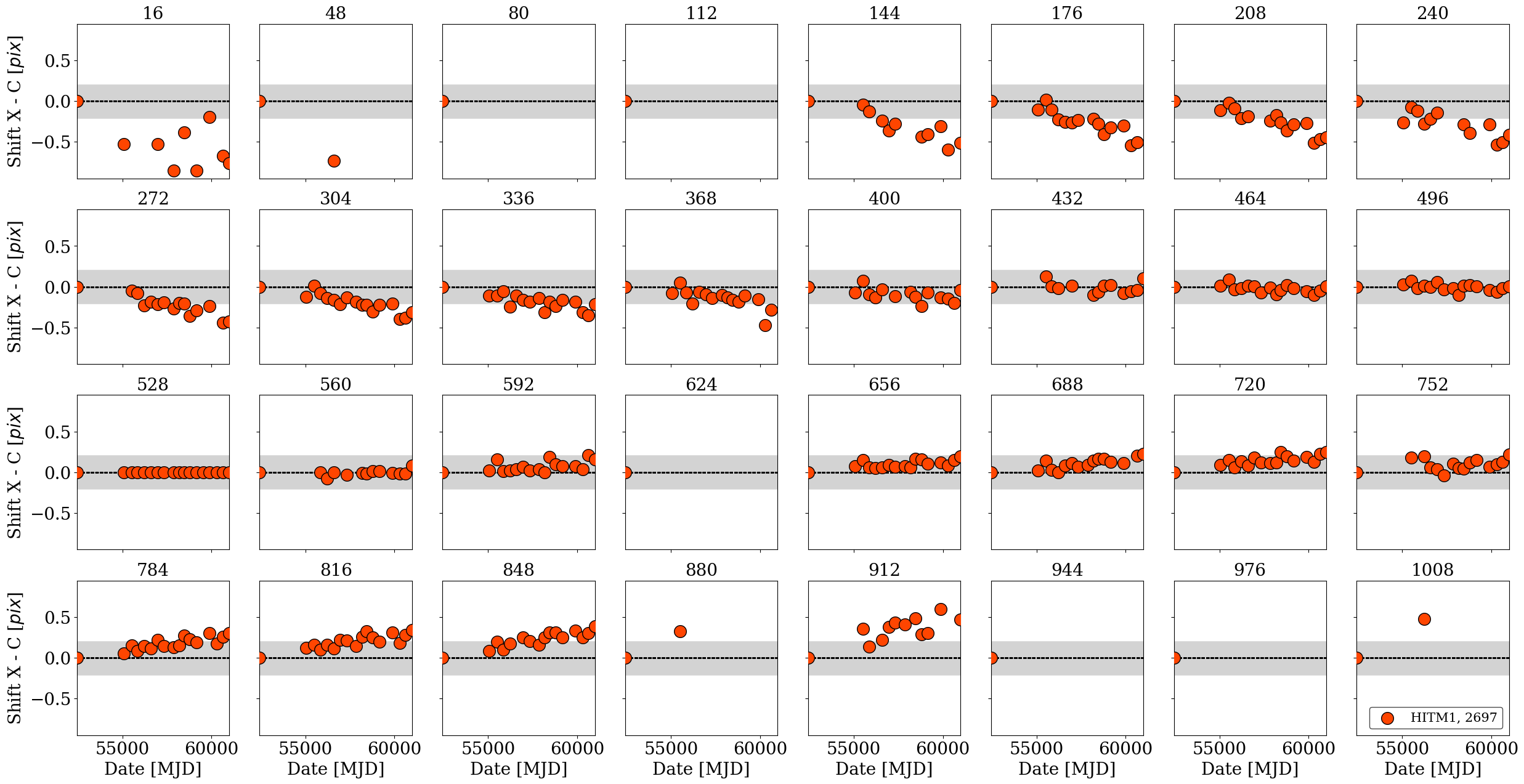}
    \includegraphics[width=.8\textwidth]{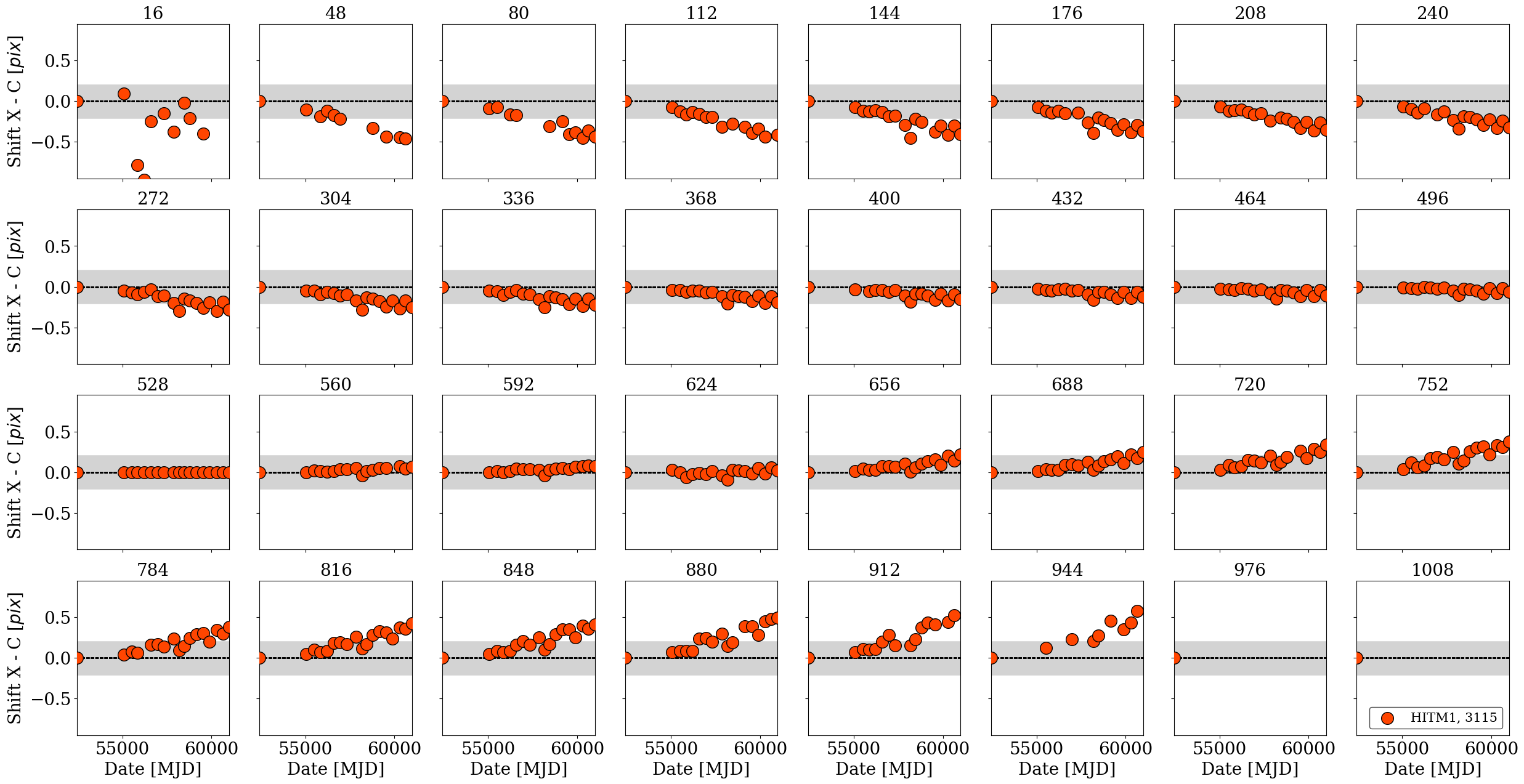}
      \caption{Same as Figure~\ref{fig:g750l-crosscorr} for G230MB/2416 (top panel), G230MB/2697 (middle panel), G230MB/3115 (bottom panel) showing the cross-correlation results obtained on the entire wavelength range. The cross-correlation returned highly scattered values, or failed entirely, at the CCD edges and, for the bluer central wavelengths, even at other detector positions due to the low signal-to-noise ratio of the extracted spectra and spurious effects (e.g., cosmic rays).}
         \label{fig:g230mb-crosscorr}
   \end{figure}

Figures~\ref{fig:g750l-crosscorr}-\ref{fig:g230mb-crosscorr} 
show the results of the cross-correlation at the 32 different positions on the CCD for G750L, G430L, G750M, G430M, G230LB and G230MB, respectively, centering the values at row 400 (closest position of our extracted spectra to the rotation center according to \citetalias{Ward-Duong2022}). 
The gray horizontal bands in the figures show the $\pm0.2$~pixels expected accuracy for the wavelength calibration.
The offsets show consistent spatial trends and comparable amplitudes (up to $\sim 1$~pixel) across all gratings, supporting the interpretation that they are dominated by CCD rotation rather than configuration-specific wavelength-solution effects.
At the edges of the CCD - note in particular row~$\sim900$ (i.e., E1 pseudo-aperture, useful to mitigate charge transfer inefficiency; see \citetalias{stisihb}, Sec.~4.2.3) - the wavelength offset goes beyond the expected accuracy in recent cycles. 
A correction to account for the wavelength shift at E1 has recently been implemented in \texttt{calstis} (see \citetalias{mingozzi2026}).
   
Figures~\ref{fig:g750m1-crosscorr}-\ref{fig:g230mb-crosscorr} also show that the cross-correlation can fail near the CCD edges, where the
extracted spectra have lower signal-to-noise ratios. This effect is
particularly evident for the G430M/4961, G230MB/2416 and G230MB/2697 settings. G230MB/1996 is also included
in the CCD dispersion-solution monitors, but the cross-correlation failed at
all extraction positions; therefore, its results are not shown here. 
This issue is discussed further in \citetalias{mingozzi2026}.
\cleardoublepage

\section{Measurements of the rotation angle and rotation rate over time}\label{sec:angles}
\begin{figure}
    \centering
      \includegraphics[width=1.\textwidth]{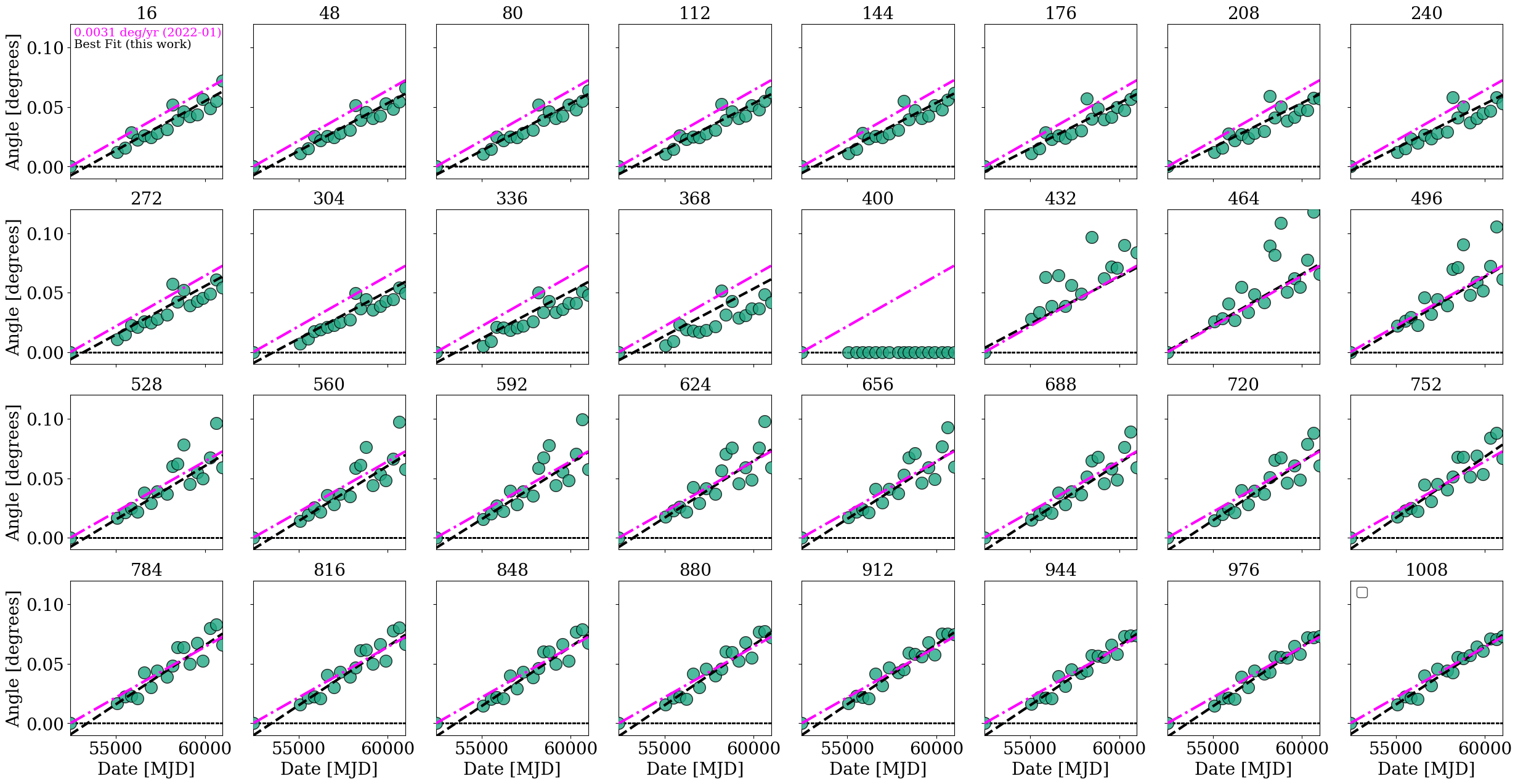}

    \includegraphics[width=1.\textwidth]{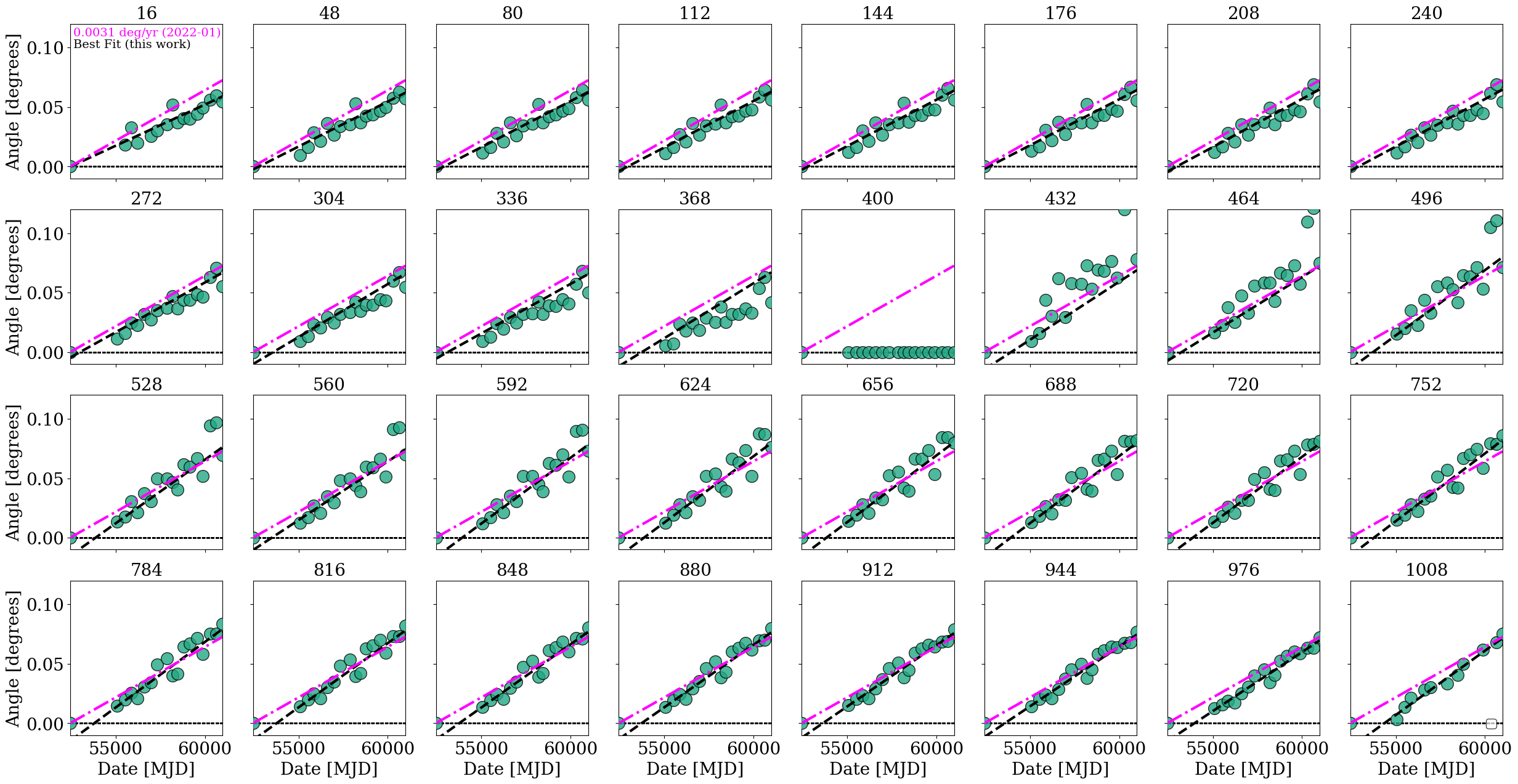}
      \caption{Top panels: Measurement of the rotation angle for G750L/7751 from the {\it $\Delta$x} values shown in Figure~\ref{fig:g750l-crosscorr}. The black dashed lines show the best-fit of the data points at each position on the CCD with a linear relation, whose slope represents the rotation rate across time. The magenta lines have the slope of $0.0031\pm0.0001$ degrees/yr obtained from \citetalias{Ward-Duong2022} - assuming the angle was 0 in Cycle 11 (dash-dotted line). Bottom panels: Same for G430L ({\it $\Delta$x} values shown in Figure~\ref{fig:g750l-crosscorr} ~\ref{fig:g430l-crosscorr}). We do not report the plots for all the other gratings, as they are analogous.}
         \label{fig:g750l-crosscorr-angle}
   \end{figure}

To measure the CCD rotation angle $\gamma$, we applied Eq.~\ref{eq:eq1} using the {\it $\Delta$x} values measured from the cross-correlation method as described in Section~\ref{sec:deltax}. 
To visualize the method, Figure~\ref{fig:g750l-crosscorr-angle} shows the derived angles at the different CCD positions for the G750L and G430L gratings.
The rotation rate over time can be measured by applying a linear fit to the scatter plot at each position on the CCD (dashed black line). 
The magenta lines show the rotation rate of $\sim0.0031$ degrees/yr from \citetalias{Ward-Duong2022}, assuming the angle was 0 degrees in Cycle 11 (dash-dotted line), used as the reference.
By definition, the angle at CCD position 400 is set to zero, as this position is adopted as the rotation center. 
Under the assumption of a rigid rotation, the angles measured at all other positions should therefore be mutually consistent (even among different gratings). We indeed obtain consistent results for all the gratings. 
The presence of scatter, and the slightly different slopes observed at the top and bottom of the CCD, may indicate that our assumptions - namely the choice of rotation center and the simplification adopted in Eq.~\ref{eq:eq1} - are not exact, although they remain reasonable. 
Over the time span from Cycle~11 to Cycle~33 (approximately two decades), the measured angles vary within the range $0$–$0.1$~degrees, which is consistent with the values reported by \citetalias{Ward-Duong2022} (see their Figure~8).

\begin{figure}
    \centering
    \includegraphics[width=.49\textwidth]{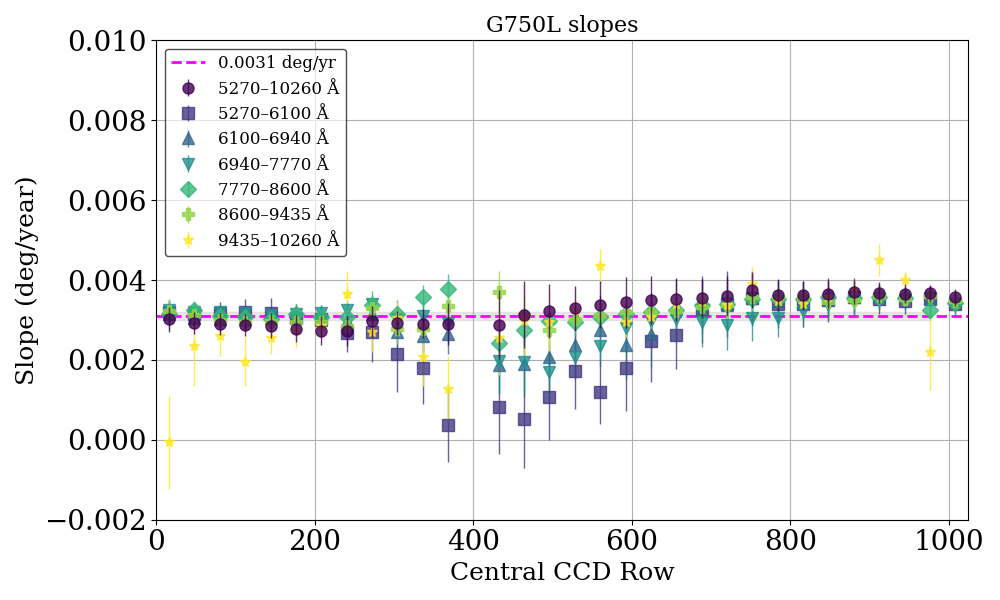}
    \includegraphics[width=.49\textwidth]{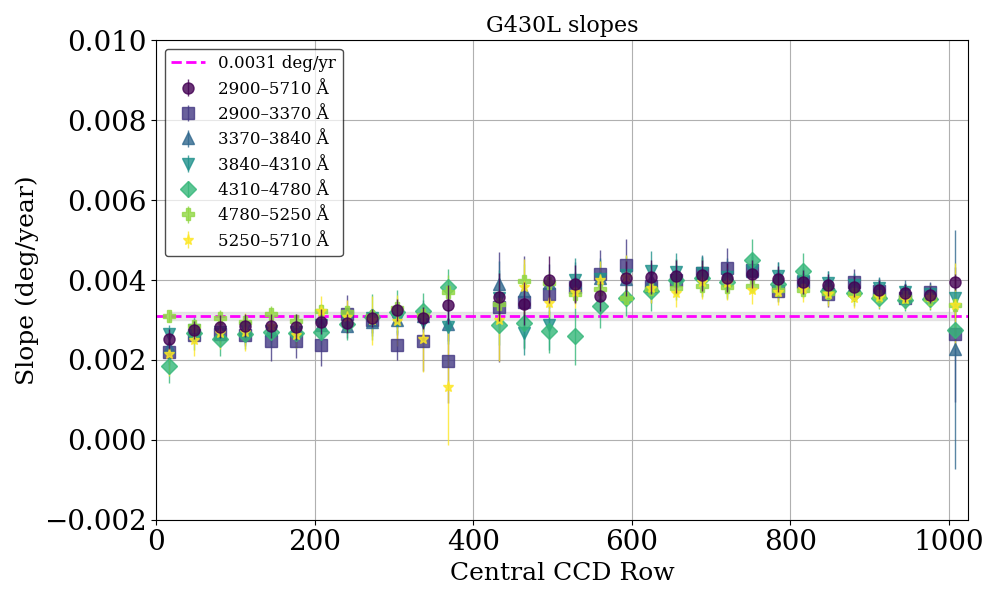}
      \caption{Best-fit slopes representing the rotation rate over years as a function of the position on the CCD for G750L/7751 (left panel) and G430L/4300 (right panel). The magenta line is the value obtained from \citetalias{Ward-Duong2022}, $0.0031\pm0.0001$~deg/yr. }
     \label{fig:lmodes-rotrates}
\end{figure}

\begin{figure}
    \centering
    \includegraphics[width=.49\textwidth]{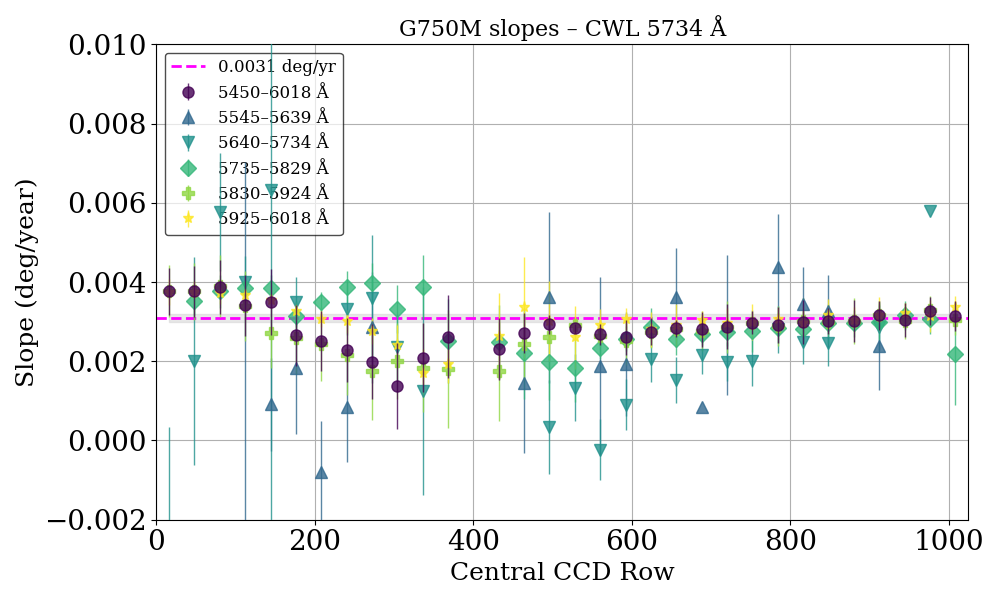}
    \includegraphics[width=.49\textwidth]{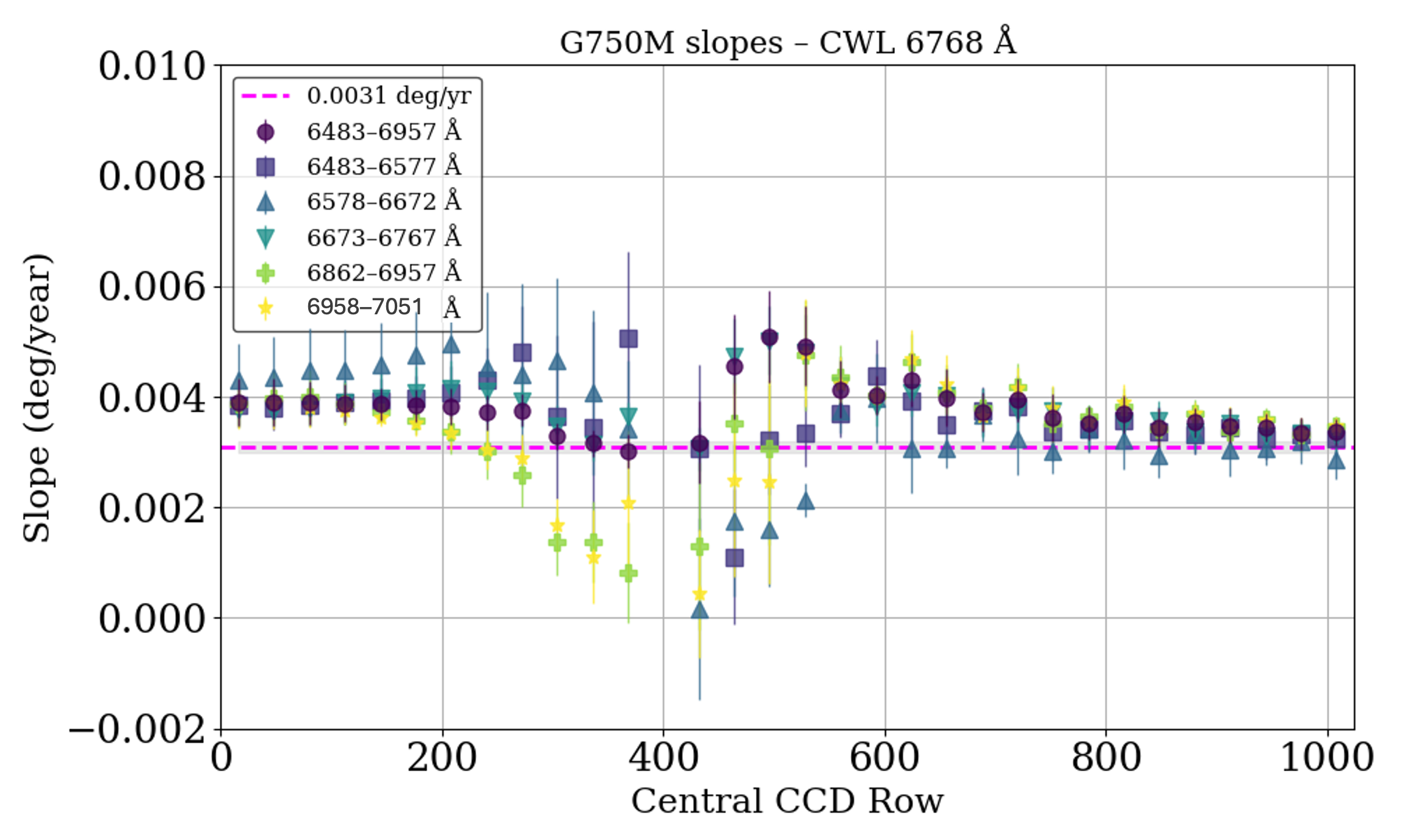}
    \includegraphics[width=.49\textwidth]{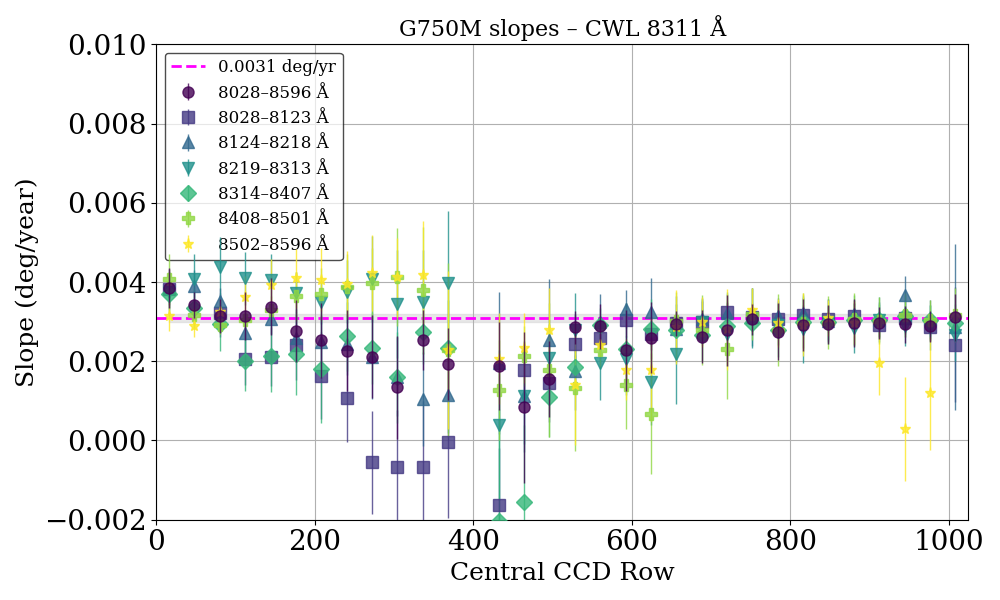}
    \includegraphics[width=.49\textwidth]{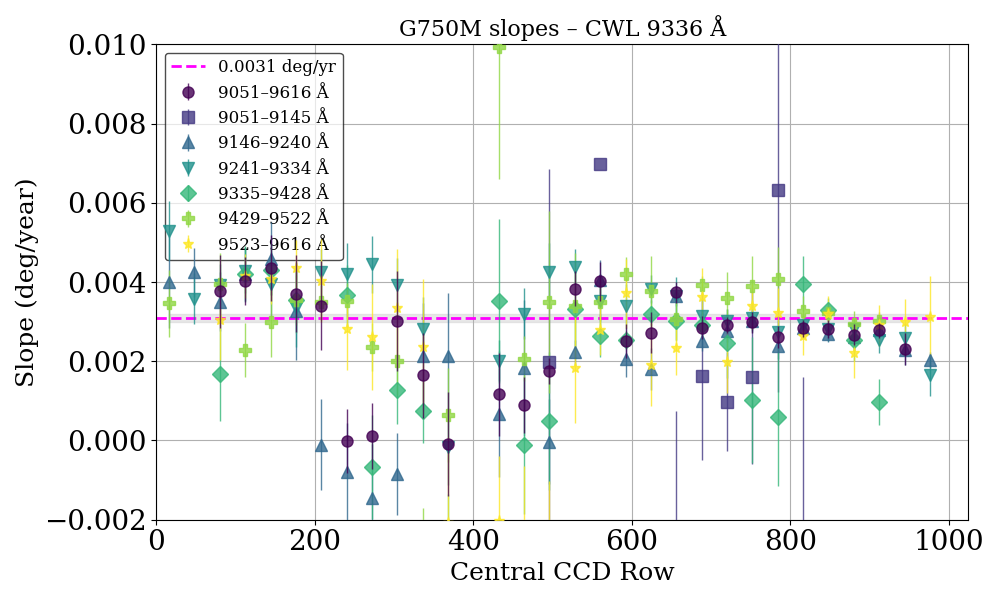}
      \caption{Same as Figure~\ref{fig:lmodes-rotrates} for G750M/5734,6768,8311,9336 central wavelengths.}
         \label{fig:g750m-rotrates}
   \end{figure}

\begin{figure}
    \centering
    \includegraphics[width=.49\textwidth]{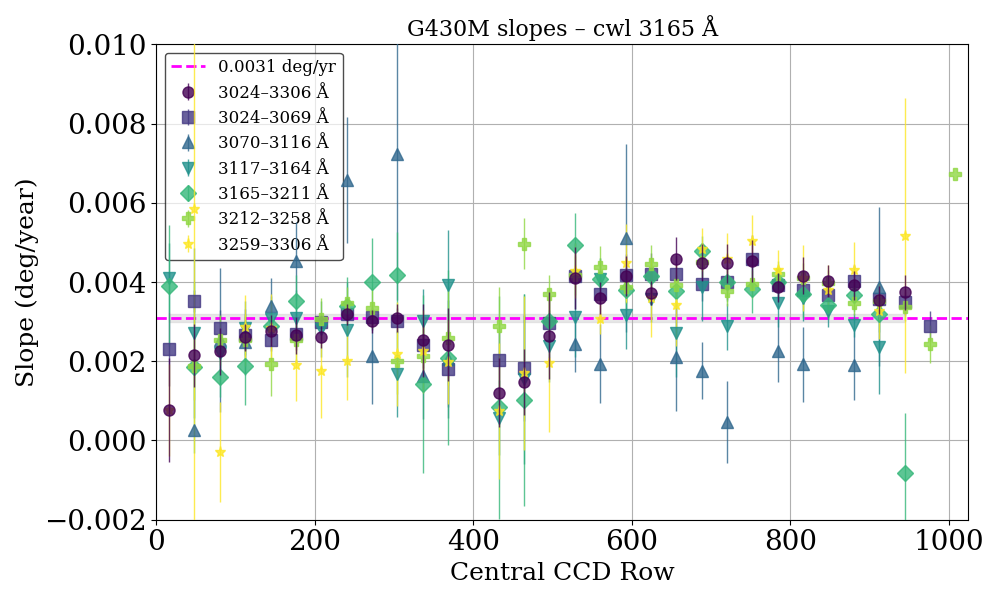}
    \includegraphics[width=.49\textwidth]{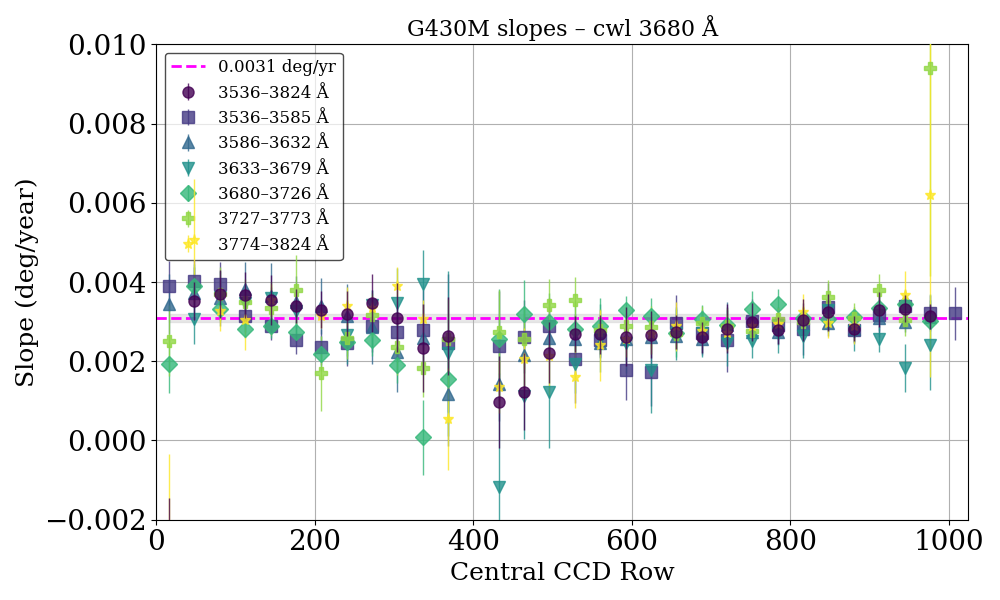}
    \includegraphics[width=.49\textwidth]{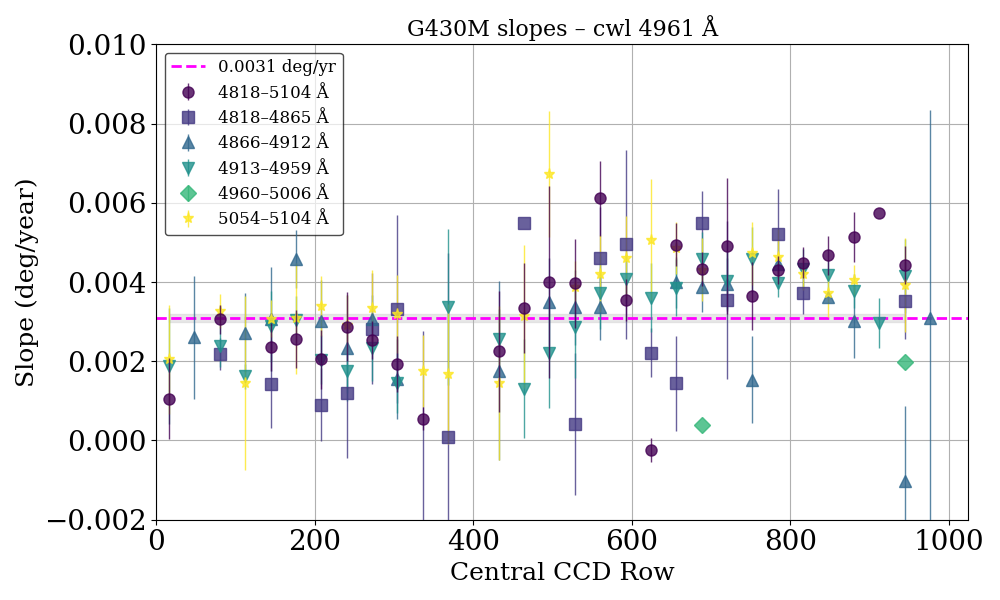}
    \includegraphics[width=.49\textwidth]{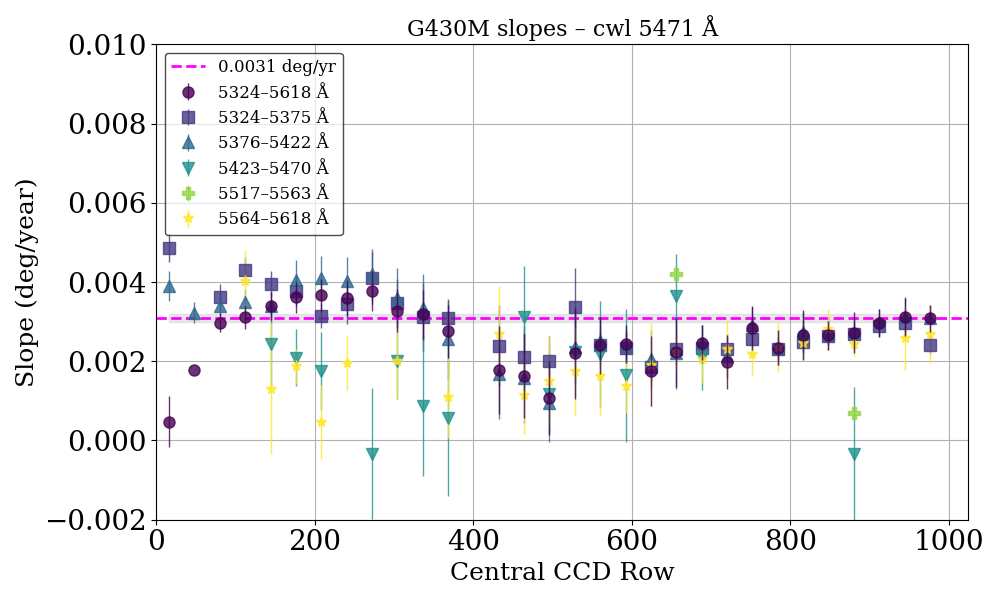}
      \caption{Same as Figure~\ref{fig:lmodes-rotrates} for G430M/3165,3680,4961,5471 central wavelengths.}
         \label{fig:g430m-rotrates}
   \end{figure}

\begin{figure}
    \centering
    \includegraphics[width=.49\textwidth]{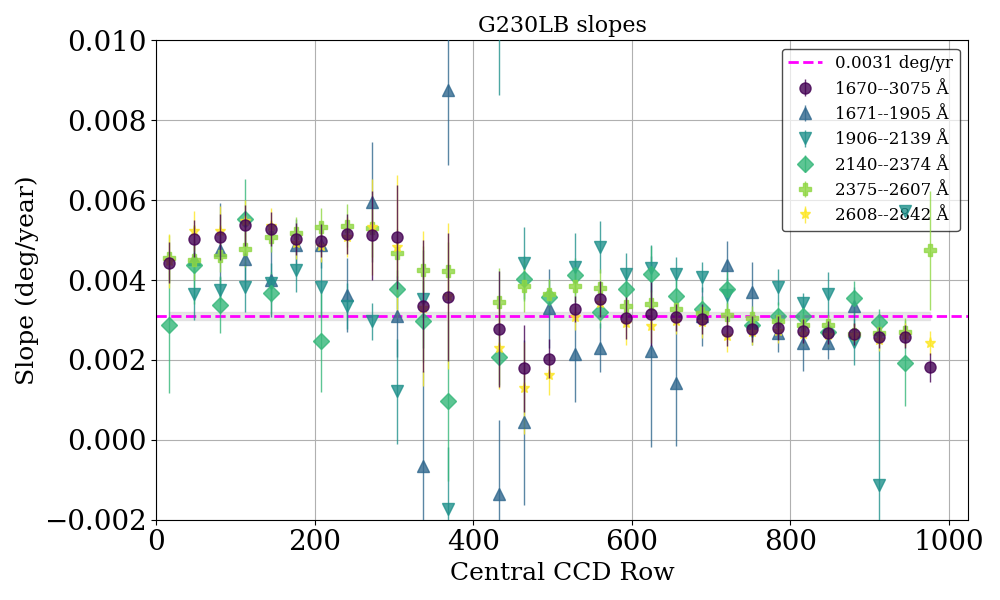}
    \includegraphics[width=.49\textwidth]{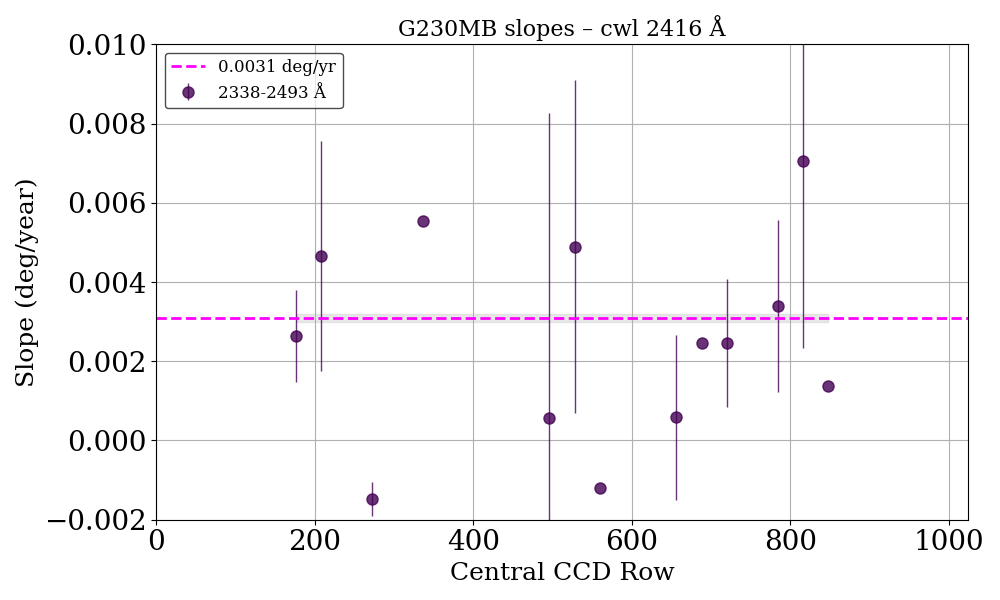}
    \includegraphics[width=.49\textwidth]{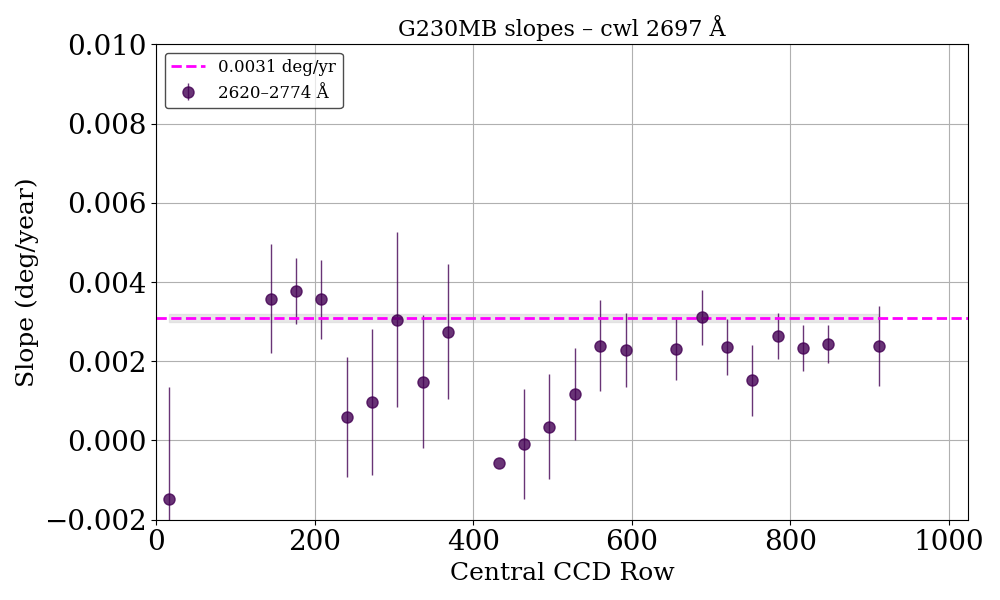}
    \includegraphics[width=.49\textwidth]{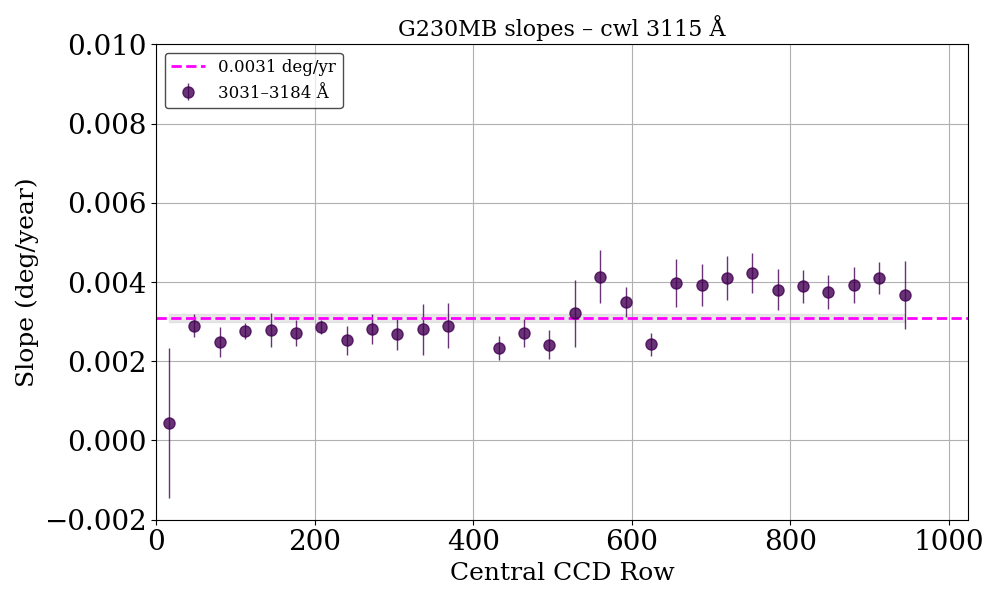}
      \caption{Same as Figure~\ref{fig:lmodes-rotrates} for G230LB/2375 and G230MB/2416,2697,3115 central wavelengths.}
         \label{fig:g230-rotrates}
   \end{figure}

The inferred rotation rates are also consistent with those reported by \citetalias{Ward-Duong2022}. We explore this further in Figures~\ref{fig:lmodes-rotrates}-\ref{fig:g230-rotrates}, which show the best-fit slopes for G750L, G430L, G750M, G430M, G230LB, and G230MB as a function of the position on the cross-dispersion axis. Different colors indicate slopes measured over different wavelength sub-ranges, where applicable.
As noted above, the scatter in the slopes may imply that our assumptions may not be fully accurate, but reasonable. 
Within 3$\sigma$, our findings are also in agreement with \citetalias{dressel2007}, which measured rotation rates for a subset of the gratings, finding rotation rates of $0.0031\pm 0.0003$ for G230LB, $0.0041\pm0.0001$ for G430L, $0.0037\pm0.0003$ for G750L, and $\sim0.0037-0.0053$ for G750M/6768,6581,8561.

As reference for future follow-ups monitoring, the median values of the measured rotation rates are reported in Tab.~\ref{tab:grating_rates_750M},~\ref{tab:grating_rates_430M}, and ~\ref{tab:grating_rates_230M} for G750, G430 and G230 configurations, respectively.
   
\begin{table}[!h]
\centering
\begin{tabular}{l c c c}
\hline
\textbf{Grating} & \textbf{Central wavelength (\AA)} & \textbf{$\Delta\lambda$ (\AA)} & \textbf{Rotation Rate (deg/yr)} \\
\hline
G750L   & 7751 & 5270--10260 & 0.0033 $\pm$ 0.0004 \\
&& 5270--6100 & 0.0031 $\pm$ 0.0006 \\
&& 6100--6940 & 0.0031 $\pm$ 0.0004 \\
&& 6940--7770 & 0.0031 $\pm$ 0.0004 \\
&& 7770--8600 & 0.0032 $\pm$ 0.0003 \\
&& 8600--9435 & 0.0032 $\pm$ 0.0003 \\
&& 9435--10260 & 0.0030 $\pm$ 0.0004 \\
\hline
G750M   & 5734 & 5450--6018 & 0.0029 $\pm$ 0.0005 \\
&& 5450--5544 & - \\
&& 5545--5639 & 0.002 $\pm$ 0.001 \\
&& 5640--5734 & 0.0023 $\pm$ 0.0007 \\
&& 5735--5829 & 0.0029 $\pm$ 0.0004 \\
&& 5830--5924 & 0.0029 $\pm$ 0.0006 \\
&& 5924--6018 & 0.0031 $\pm$ 0.0004 \\
\hline
G750M   & 6768 & 6483--7051 & 0.0038 $\pm$ 0.0003 \\
&& 6483--6577 & 0.0036 $\pm$ 0.0004 \\
&& 6578--6672 & 0.0034 $\pm$ 0.0007 \\
&& 6673--6767 & 0.0038 $\pm$ 0.0004 \\
&& 6768--6861 & - \\
&& 6862--6957 & 0.0036 $\pm$ 0.0004 \\
&& 6958--7051 & 0.0036 $\pm$ 0.0004 \\
\hline
G750M   & 8311 & 8028--8596& 0.0029 $\pm$ 0.0007 \\
&& 8028--8123 & 0.0026 $\pm$ 0.0006 \\
&& 8124--8218 & 0.0030 $\pm$ 0.0006 \\
&& 8219--8313 & 0.0030 $\pm$ 0.0008 \\
&& 8314--8407 & 0.0028 $\pm$ 0.0007 \\
&& 8408--8501 & 0.0030 $\pm$ 0.0007 \\
&& 8502--8596 & 0.0029 $\pm$ 0.0008 \\
\hline
G750M   & 9336 & 9051--9616& 0.0028 $\pm$ 0.0005 \\
&& 9051--9145 & - \\
&& 9146--9240 & 0.0024 $\pm$ 0.0006 \\
&& 9240--9334 & 0.0024 $\pm$ 0.0005 \\
&& 9335--9428 & 0.0025 $\pm$ 0.0008 \\
&& 9429--9522 & 0.0034 $\pm$ 0.0007 \\
&& 9523--9616 & 0.0028 $\pm$ 0.0009 \\
\hline
\end{tabular}
\caption{Median rotation rates for G750L/7751 and G750M/5734,6768,8311,9336. The values were derived by calculating the {\it $\Delta$x} shifts in pixels through cross-correlation of spectra at 32 different CCD positions, using Cycle 11 as a reference. Cross-correlation was performed across the full wavelength range of each grating and within six sub-ranges to verify that shifts were not significantly wavelength-dependent. The angular shift, computed from {\it $\Delta$x}, is plotted versus time, and the rotation rate (deg/yr) is obtained as the slope of the linear fit to these data.}
\label{tab:grating_rates_750M}
\end{table}

\begin{table}[!h]
\centering
\begin{tabular}{l c c c}
\hline
\textbf{Grating} & \textbf{Central wavelength (\AA)} & \textbf{$\Delta\lambda$ (\AA)} & \textbf{Rotation Rate (deg/yr)} \\
\hline
G430L   & 4300 & 2900--5710  & 0.0036 $\pm$ 0.0003 \\
&& 2900--3370 & 0.0035 $\pm$ 0.0004 \\
&& 3370--3840 & 0.0038 $\pm$ 0.0004 \\
&& 3840--4310 & 0.0036 $\pm$ 0.0003 \\
&& 4310--4780 & 0.0032 $\pm$ 0.0004 \\
&& 4780--5250 & 0.0036 $\pm$ 0.0002 \\
&& 5250--5710 & 0.0035 $\pm$ 0.0004 \\
\hline
G430M   & 3165 & 3024--3306 & 0.0032 $\pm$ 0.0005 \\
&& 3024--3069 & 0.0033 $\pm$ 0.0004 \\
&& 3070--3116 & 0.003 $\pm$ 0.001 \\
&& 3117--3164 & 0.0030 $\pm$ 0.0008 \\
&& 3165--3211 & 0.0036 $\pm$ 0.0007 \\
&& 3212--3258 & 0.0035 $\pm$ 0.0005 \\
&& 3259--3306 & 0.003 $\pm$ 0.001 \\  
\hline
G430M   & 3680 & 3536--3824& 0.0029 $\pm$ 0.0005 \\
&& 3536--3585 & 0.0028 $\pm$ 0.0006 \\
&& 3586--3632 & 0.0028 $\pm$ 0.0006 \\
&& 3633--3679 & 0.0027 $\pm$ 0.0006 \\
&& 3680--4726 & 0.0029 $\pm$ 0.0004 \\
&& 3727--3773 & 0.0030 $\pm$ 0.0006 \\
&& 3774--3824 & 0.0029 $\pm$ 0.0005 \\
\hline
G430M   & 4961 & 4818--5104 & 0.0036 $\pm$ 0.0007 \\
&& 4818--4865 & 0.003 $\pm$ 0.001 \\
&& 4866--4912 & 0.003 $\pm$ 0.001 \\
&& 4913--4959 & 0.0031 $\pm$ 0.0007 \\
&& 4960--5006 & - \\
&& 5007--5053 & - \\
&& 5054--5104 & 0.0035 $\pm$ 0.0009 \\
\hline
G430M   & 5471 & 5324--5618 & 0.0027 $\pm$ 0.0005 \\
&& 5324--5375 & 0.0027 $\pm$ 0.0004 \\
&& 5376--5422 & 0.0028 $\pm$ 0.0005 \\
&& 5423--5470 & 0.002 $\pm$ 0.001 \\
&& 5470--5516 & - \\
&& 5517--5563 & - \\
&& 5564--5618 & 0.0020 $\pm$ 0.0008 \\
\hline
\end{tabular}
\caption{Same as Tab.~\ref{tab:grating_rates_750M} for G430L/4300 and G430M/3165,3680,4961,5471.}
\label{tab:grating_rates_430M}
\end{table}

\begin{table}[!h]
\centering
\begin{tabular}{l c c c}
\hline
\textbf{Grating} & \textbf{Central wavelength (\AA)} & \textbf{$\Delta\lambda$ (\AA)} & \textbf{Rotation Rate (deg/yr)} \\
\hline
G230LB & 2375 & 1671--3075 & 0.0031 $\pm$ 0.0004 \\
&& 1671--1905 & - \\
&& 1906--2139 & 0.0033 $\pm$ 0.0009 \\
&& 2140--2374 & 0.0038 $\pm$ 0.0006 \\
&& 2375--2607 & 0.0033 $\pm$ 0.0005 \\
&& 2608--2842 & 0.0038 $\pm$ 0.0004 \\
&& 2843--3075 & 0.0030 $\pm$ 0.0004 \\  
\hline
G230MB   & 2416 & 2338--2493 & - \\
\hline
G230MB   & 2697 & 2620--2774 & 0.002 $\pm$ 0.001 \\
\hline
G230MB   & 3115 & 3031--3184 & 0.0029 $\pm$ 0.0004 \\
\hline
\end{tabular}
\caption{Same as Tab.~\ref{tab:grating_rates_750M} for G230LB/2375 and G230MB/2416,2697,3315.}
\label{tab:grating_rates_230M}
\end{table}

\clearpage

\section{Conclusions}\label{sec:conclusion}
In this ISR, we presented the rotation angle rates measured from the STIS CCD G750L/7751, G750M/5734, 6768, 8311, 9336, G430L/4300, G430M/3165, 3680, 4961, 5471, G230LB/2375 and G230MB/2416, 2697, 3315 gratings. 
The median values of the measured rotation rates are reported in Tab.~\ref{tab:grating_rates_750M},~\ref{tab:grating_rates_430M} and ~\ref{tab:grating_rates_230M}.
Overall, these rotation rates are consistent within $\sim$1-3$\sigma$ with those measured by \citetalias{Ward-Duong2022} and \citetalias{dressel2007}.
We highlight that our measurements can be biased by different assumptions we made: 1) A rigid rotation of the CCD; 2) The assumption of the rotation center at row $\sim400$ from \citetalias{Ward-Duong2022}; 3) The simplification of Eq.~\ref{eq:eq1}. However, the agreement with previous works using completely different approaches is reassuring.

Overall, we confirmed that CCD rotation affects the wavelength calibration accuracy, which in the most recent cycles no longer remains within the expected $\pm0.2$~pixel level near the edges of the detector, as shown in Figures~\ref{fig:g750l-crosscorr}-~\ref{fig:g230mb-crosscorr} (see also \citetalias{welty2018}). 
A parallel investigation by the STIS team, presented in \citetalias{mingozzi2026}, describes a correction implemented in the \texttt{calstis} pipeline to improve the wavelength calibration accuracy at the E1 position (row~$900$, near the upper edge of the CCD). It also provides a notebook for evaluating, and when needed correcting, the wavelength calibration accuracy at arbitrary positions across the detector. 
The STIS team continues to monitor the CCD rotation and its effects on both the observed data and the calibration pipeline.


\vspace{-0.3cm}
\ssectionstar{Acknowledgements}
\vspace{-0.3cm}
We thank Emily Rickman for her constructive review of the initial version of this ISR, and Joleen Carlberg and Daniel Welty for helpful discussions during the development of this project.

\ssectionstar{Change History for STIS ISR 2026-03}\label{sec:History}
\vspace{-0.3cm}
Version 1: \ddmonthyyyy\today - Original Document 

\setlength{\bibsep}{0in}
\bibliography{mybib}

\begin{thebibliography}{8}
\providecommand{\natexlab}[1]{#1}
\providecommand{\url}[1]{\texttt{#1}}
\expandafter\ifx\csname urlstyle\endcsname\relax
  \providecommand{\doi}[1]{doi: #1}\else
  \providecommand{\doi}{doi: \begingroup \urlstyle{rm}\Url}\fi

\bibitem[{Dressel} et~al.(2007){Dressel}, {Bohlin}, {Lindler}, and
  {Holfeltz}]{dressel2007}
L.~{Dressel}, R.~{Bohlin}, D.~{Lindler}, and S.~{Holfeltz}.
\newblock {Time Dependent Trace Angles for the STIS First Order Modes}.
\newblock STIS ISR 2007-03, August 2007.

\bibitem[{Ward-Duong} et~al.(2022){Ward-Duong}, {Lockwood}, {Debes}, and {De
  Rosa}]{Ward-Duong2022}
K.~{Ward-Duong}, S.~{Lockwood}, J.~{Debes}, and R.~J. {De Rosa}.
\newblock {Long-Term Rotational Evolution of the STIS CCD Flatfields}.
\newblock STIS ISR 2022-01, March 2022.

\bibitem[{Nguyen} et~al.(2021){Nguyen}, {De Rosa}, and {Kalas}]{nguyen2021}
M.~M. {Nguyen}, R.~J. {De Rosa}, and P.~{Kalas}.
\newblock {First Detection of Orbital Motion for HD 106906 b: A Wide-separation
  Exoplanet on a Planet Nine-like Orbit}.
\newblock \emph{\aj}, 161\penalty0 (1):\penalty0 22, January 2021.
\newblock \doi{10.3847/1538-3881/abc012}.

\bibitem[{Friedman}(2005)]{friedman2005}
S.~D. {Friedman}.
\newblock {Wavelength Calibration Accuracy of the First-Order CCD Modes Using
  the E1 Aperture}.
\newblock STIS ISR 2005-03, August 2005.

\bibitem[{Welty}(2018)]{welty2018}
D.~{Welty}.
\newblock {Monitoring the STIS Wavelength Calibration: MAMA and CCD First-Order
  Modes}.
\newblock STIS ISR 2018-04, August 2018.

\bibitem[{Welty} et~al.(2025){Welty}, {Bohlin}, {Carlberg}, {Dallas},
  {Hernandez}, {Jones}, {Lockwood}, {Medallon}, {Rickman}, {Stapleton}, and
  {Wheeler}]{welty2025}
D.~{Welty}, R.~{Bohlin}, J.~{Carlberg}, M.~{Dallas}, S.~{Hernandez},
  A.~{Jones}, S.~{Lockwood}, S.~{Medallon}, E.~{Rickman}, D.~{Stapleton}, and
  T.~{Wheeler}.
\newblock {STIS Cycle 30 Calibration Programs}.
\newblock STIS ISR 2025-01, January 2025.

\bibitem[{Mingozzi} et~al.(2026){Mingozzi}, {Jedrzejewski}, {Siebert}, {Welty},
  {Lockwood}, and {Carlberg}]{mingozzi2026}
M.~{Mingozzi}, R.~{Jedrzejewski}, M.~{Siebert}, D.~{Welty}, S.~{Lockwood}, and
  J.~{Carlberg}.
\newblock {Wavelength Calibration Accuracy Across the STIS CCD: Pipeline Update
  and User Guidance}.
\newblock STIS ISR 2026-04, August 2026.

\bibitem[{Rickman} and {Brown}(2025)]{stisihb}
Emily {Rickman} and Jacqueline {Brown}.
\newblock {Space Telescope Imaging Spectrograph Instrument Handbook for Cycle
  34, v.25}.
\newblock 2025.

\end{thebibliography}

\newpage
\appendix

\vspace{-0.3cm}
\setcounter{section}{0}
\renewcommand{\thesection}{\Alph{section}}

\appsection{Rotation Angle Computation}\label{sec:Appendix}

For a point $P_1=(x_1,y_1)$ rigidly rotated around a center 
$(x_0,y_0)$ to a final position $P_1'=(x_1',y_1')$, the absolute value of the rotation angle is given by:

\begin{equation}
|\gamma|=
\arccos
\left[
\frac{
(x_1-x_0)(x_1'-x_0)+(y_1-y_0)(y_1'-y_0)
}{
r^2
}\right],
\label{eq:gamma_scalar}
\end{equation}

where

\begin{equation}
r=\sqrt{(x_1-x_0)^2+(y_1-y_0)^2}.
\label{eq:r_definition}
\end{equation}

\noindent In the specific case considered in this ISR, the rotation angle is small and the
resulting displacement can be measured directly from detector coordinate shifts.
It is therefore convenient to derive a simpler expression for $\gamma$.

\medskip
\noindent
\textbf{Exact coordinate transformation under rotation}

A rotation of a point around the center $(x_0,y_0)$ by an angle $\gamma$
transforms the coordinates according to

\begin{align}
x_1'-x_0 &= (x_1-x_0)\cos\gamma-(y_1-y_0)\sin\gamma
\label{eq:rotation_x}\\
y_1'-y_0 &= (x_1-x_0)\sin\gamma+(y_1-y_0)\cos\gamma
\label{eq:rotation_y}
\end{align}

Define the coordinate shifts as

\begin{equation}
\Delta x=x_1'-x_1
\qquad
\Delta y=y_1'-y_1
\label{eq:shift_definition}
\end{equation}

Substituting Equations~\ref{eq:rotation_x} and \ref{eq:rotation_y}
into Equation~\ref{eq:shift_definition} gives

\begin{align}
\Delta x &=(x_1-x_0)(\cos\gamma-1)-(y_1-y_0)\sin\gamma
\label{eq:dx_exact}\\
\Delta y &=(x_1-x_0)\sin\gamma+(y_1-y_0)(\cos\gamma-1)
\label{eq:dy_exact}
\end{align}

These expressions describe the exact displacement produced by a rotation
of angle $\gamma$ around the center $(x_0,y_0)$.

\medskip
\noindent
\textbf{Small-angle approximation}

According to \citetalias{Ward-Duong2022}'s rotation rate measurement, the rotation between Cycle~11 and
Cycle~33 is clockwise ($\gamma<0$ in the convention adopted here) with

\[
|\gamma|
\approx
0.0031\,\mathrm{deg\,yr^{-1}}\times29\,\mathrm{yr}
\approx
0.0899\,\mathrm{deg}
\approx
0.001569\,\mathrm{rad}
\]

Since $|\gamma| \ll 1$, the small-angle approximations
\begin{equation}
\sin\gamma\approx\gamma
\qquad
\cos\gamma\approx1
\label{eq:small_angle}
\end{equation}
are justified in the derivation below.

Substituting Equation~\ref{eq:small_angle} into
Equations~\ref{eq:dx_exact} and \ref{eq:dy_exact} gives

\begin{align}
\Delta x &\approx-\gamma(y_1-y_0)
\label{eq:dx_small}\\
\Delta y &\approx\phantom{-}\gamma(x_1-x_0)
\label{eq:dy_small}
\end{align}

From Equation~\ref{eq:dx_small}, the rotation angle can be estimated directly from the measured horizontal displacement and vertical distance from the rotation center as

\begin{equation}
\gamma \approx -\frac{\Delta x}{y_1-y_0}
\label{eq:gamma_from_dx}
\end{equation}

Taking the absolute value gives

\begin{equation}
|\gamma| \approx \frac{|\Delta x|}{|y_1-y_0|}
\label{eq:gamma_from_dx_abs}
\end{equation}

\end{document}